# A mathematical and Physical Model Improves Accuracy in Simulating Material Relaxation Modulus and Viscoelastic Responses

*Qinwu Xu[1,*], Björn Engquist[1,2,†]*
*1. Institute for Computational Engineering and Sciences (ICES); 2. Department of Mathematics*
*University Station, the University of Texas at Austin, Austin, TX 78712 USA*
** qinwu.xu@utexas.edu, †engquist@ices.utexas.edu*

we propose a new material viscoelastic model and mathematical solution to simulate relaxation modulus $E(t)$ and viscoelastic response. The model formula of $E(t)$ is extended from sigmoidal function considering nonlinear strain hardening and softening. Its physical mechanism can be interpreted by a spring network-viscous medium model with only five parameters in a simpler format than the molecular-chain based polymer models to represent general materials. We also developed a three-dimensional finite-element method and robust numerical algorithms to implement this model for solving partial differential equations. We validate the model through both experimental data and numerical simulations on a broad range of materials including bitumen, shape-memory polymer, spider-inspired silk, hydrogel, biomaterials and bone. By satisfying the 2nd law of thermodynamics in the form of Calusius-Duhem inequality, the model is able to simulate creep and sinusoidal deformation, and energy dissipation. As compared to Prony series – the most general model being used often with a large number of model parameters, the proposed model has improved accuracy in fitting experimental data and predicting $E(t)$outside of the experimental range, and the latter one is especially useful for material design. The new model also has higher numerical accuracy while competitive numerical stability and computation speed for convergence.

## I. INTRODUCTION

Viscoelastic material exhibits both elastic and viscous characteristics, and its deformation is temperature dependent with thermal transitions. Most materials including the advanced and bio materials somehow exhibit viscoelastic behavior, including nanotube composites[1], shape memory polymer[2], dielectric elastomers[3,4], hydrogels[5], natural and synthetic spider silks[6], bioinspired metal-crosslinked polymer[7], and bones[8]. It is important to accurately simulate the stress behavior and viscoelastic deformation of the subjects in order to evaluate and design advanced materials. Thermal transitions of viscoelastic materials can be described either in terms of free volume changes or relaxation time[9]. Relaxation modulus $E(t)$ is a characteristic of material viscoelasticity as used to describe the stress relaxation of materials with time. Different theories and models for describing $E(t)$ were proposed primarily in the field of polymer science. The Crankshaft model[10] simulates polymer molecular as a series of jointed segments involving in a few stages of thermal transition with elevated time or temperature. For example, with the increase of relaxation time, the first or $\gamma$ transition may start at the local motions of molecular, and then possibly $\beta$ transition appears in which $E(t)$ drops slightly due to the bend and stretch of molecules with elevated temperature, consequently the glass transition occurs in which $E(t)$ significantly reduces until reaching the rubbery stage, and lastly the terminal transition may exit when polymer melts into liquids due to the slippage of molecular chains. Different materials may present different thermal transition behaviors, but generally all include the glass transition at the solid state, which is also the focus of this study.

Physical models have been developed to describe linear viscoelasticity. Based on the molecular dynamics theory, the Rouse model simulates the single chain diffusion of polymers as Brownian motion of the beads–and harmonic springs system[11] (see **Figure 1**a). The Kremer-Grest model used up to hundreds of chains and beads in its simulation work[12]. The well-known tube model and theory described that entangled polymer chains are confined in tubes with permanent topological interactions and move along tubes[13,14] (see Figure 1b). The stress relaxation of each chain is calculated as the fraction of the tube that has not been vacated. The arm reptation model was also proposed[14,15], in which the entangled monomers (unit of polymer macromolecular) are retracted by arms (see Figure 1c).

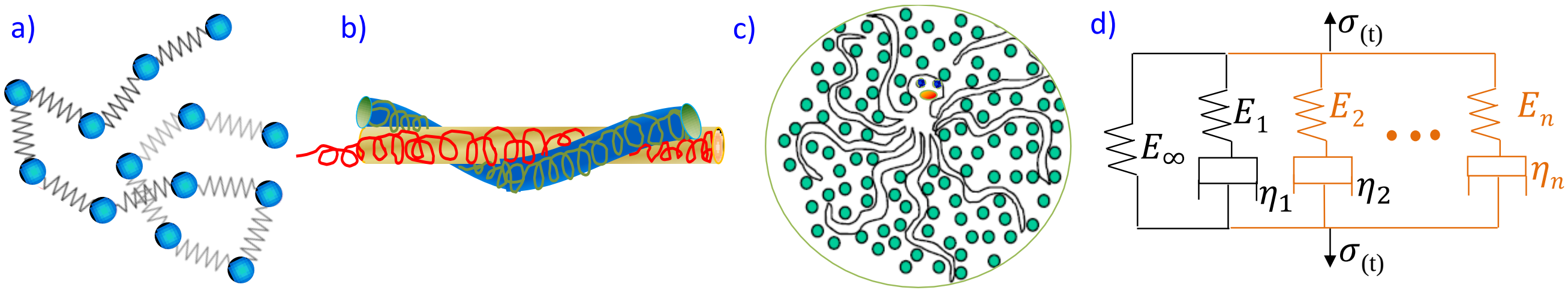

**Figure 1**. Material viscoelastic models: a) friction beads-spring model; b) tube model; c) arm reptation model and d) generalized Maxwell model.

Among the physical models the Maxwell and generalized Maxwell models are the most general ones being used to simulate the glass transition of materials and fit experimental data. Maxwell model proposed after Maxwell and Wiechert in late 19th century consists of a linear spring and a dashpot in series. With an additional spring $E_\infty$ in Parallel it is known as the standard linear solid or Zener model.

For the viscosity of the dashpot the stress is linearly depenent on the strain rate such that $\sigma(t) = \eta\dot{\varepsilon}(t)$ with $\eta$ as the viscosity parameter. However, the Maxwell model can only capture the relaxation behavior of polymers in a very limited time range. Accordingly, a series of such models have been assembled in parallel, delivering a physical system referred as the generalized Maxwell (GM) model as shown in Figure 1**d**. The combination of elastic springs and viscous dashpots of this physical system was interpreted as different molecular chain segment lengths under different time distributions for polymer structures[16]. The generalized Maxwell model expresses $E(t)$ as follows:

$$E(t) = E_\infty + \sum_{i=1}^{n} E_i e^{-\frac{E_i}{\eta_i}t} = E_\infty + \sum_{i=1}^{n} E_i e^{-\ t/\tau_i} \quad (1)$$

where $E_\infty$ is the model's elastic modulus at infintie time $t = \infty$, $E_i$ is the elastic modulus of the spring, $\eta_i$ is the viscosity of dashpot, $n$ is the number of spring-dashpot series, and $\tau_i = \frac{\eta_i}{E_i}$ is the relaxation time spectrum representing a time range for modulus decay from $E_i$ to 0 for a single spring-dashpot series.

With accumulated spring-dashpot chains in parallel the $E(t)$ of the system is the sum of $E_i$ distributed at variable time spectrums. This formula with accumulated exponential terms in a discrete spectrum with finite number of series is also named as Prony series (PS). Each $E_i$ and $\eta_i$ in PS are constant, and a combination of multiple $E_i$ and $\eta_i$ significantly improves numerical accuracy in fitting experimental data as compared to that of the standard Zener model. The model parameters are determined by fitting on experimental data using optimization skills. Usually a higher $n$ value may achieve a higher fitting accuracy, but also involves more complexity and variability, which will be further discussed later.

PS has been adopted as the most general model for most materials to fit on experimental data and simulate stress relaxations within the glass transition for modeling linear viscoelasticity. Examples include conventional materials of polymers[17,18], glasses[19], silicon[20], asphalt concrete (AC)[21], and multi-functional materials of dielectric elastomers[34], shape-memory polymers[2], bioinspired metal-crosslinked polymer[7], and bio-/tissue materials of ligament[22], skin[23], brain[24], and blood vessel[25]. PS in exponential format is computationally convenient for the recursive integration[26]. It has also been implemented as a standard material input for simulating responses of materials and structures in the powerful multiphysics software (e.g. ANSYS and ABAQUS).

There are other models proposed to simulate specific materials or behaviors. Nguyen et al.[27] proposed a thermoviscoelastic model for simulating stress relaxation mechanism of a family of (meth) acrylate-based networks of amorphous shape memory polymers. Sakai et al proposed an empirical model to simulate viscoelasticity of bones based on two Kohlrausch–Williams–Watts functions as follows[28]:

$$E(t) = E_0\left[A_1 e^{-\left(\frac{t}{\tau_1}\right)^{\beta}} + (1 - A_1)e^{-\left(\frac{t}{\tau_2}\right)^{\beta}}\right] \tag{2}$$

The fractional derivative model has also been proposed by changing the viscosity function as a fractional order of time such that[29]:

$$\sigma_{\eta}(t) = \frac{\eta d\varepsilon^{\alpha}(t)}{dt^{\alpha}} \tag{3}$$

where $\alpha$ is a parameter that $N < \alpha < N + 1$ ($N$ is an integer).

The stress-strain relationship of the viscoelastic system using fractional derivative model is expressed as follows:

$$\sigma(t) = E_0\varepsilon(t) + E_1 D^{\alpha}[\varepsilon(t)] \tag{4}$$

where $D$ is a fractional derivative, $E_0$ and $E_1$ are two modulus parameters.

The fractional derivative model can reduce the number of parameters while sustaining desired accuracy, although it involves more complexity in fitting experimental data and simulating responses of materials and structures. The Huet-Sayegh model uses only five model parameters to express material dynamic modulus in fractional order[31]. However, it is unable to simulate the stress relaxation and deformation within time domain[32].

Nonlinearity of viscoelastic models are also considered for materials with large deformation[33] or properties changing with deformation. Here we just present some examples: it may consider the spring's elastic modulus as a nonlinear function of time (e.g. Schapery's model[34]), or the stress of dashpot as a nonlinear function of strain rate (e.g., the power-law[35] and exponential function[36]). However, computation instability may arise as a mathematical concern when using nonlinear models[3738].

In summary, a few research questions may arise for existing approaches and PS model. The multi-molecular-chain based models and theories discussed above (e.g. beads-springs, entangled tubes, and arm-starts) are more suitable for polymers or rubber-like materials than others with different morphology (e.g. amorphous asphalt and crystal metals). These models also involve difficulty or variability in accurately characterizing model parameters and numerical simulation responses for the macro- structural and material systems. Secondly or more importantly, PS– the most general model being used in fitting experimental data and simulating responses of materials and structures, can raise a few research questions. PS model is primarily used for modeling linear viscoelasticity; however, many advanced and multifunctional materials may perform nonlinear behaviors such as strain hardening or softening at relatively higher strain range. It has been noticed that PS may produce unsatisfied curve fitting on the experimental data[3239] since the formula is based on the discrete spectrum. This is especially true when a relatively small number of terms is used as often is the practice (e.g. $n$=7 or even less[40,41]). When using a large number of terms, the accuracy may be improved. However, it becomes more difficult or variable to fit a large number of model parameters mathematically although multiple optimization skills exist[39]. It also becomes difficult to provide a clear physical interpretation of this extensive spring-dashpot systems other than improving the data fitting mathematically using extensive model parameters for materials other than the multi-molecular-chain based polymers. In addition, PS model is primarily used for fitting on experimental data rather than prediction, while the predicted modulus is an essential input for simulating responses outside of experimental range toward evaluation and design of materials. Strain hardening or softening may appear for viscoelastic materials[30] especially the advanced and biomaterials, which was not taken into accounted in many existing nonlinear viscoelastic models, e.g. those mentioned above[34,35,36].

Accordingly, the objective of this research is to develop an alternative and continuous-time spectrum based viscoelastic model in simpler format for simulating the stress behavior and viscoelastic deformation of general materials to improve accuracy in experimental data fitting and prediction and numerical modeling. The model is able to consider both linear and nonlinear behaviors such as the strain hardening of advanced and bio- materials. We tested model by both experimental data and numerical simulations for a broad range of materials including infrastructure, shape-memory polymer, hydrogels, bio- and spider silk-inspired materials, and bone. We also developed a finite-element method and robust numerical algorithms to implement this model for simulating responses of materials and structures. We report the numerical accuracy, stability, and convergence of the proposed model as compared to that of PS.

## II. MODEL DEVELOPMENT

Beside the discrete-spectrum based models like PS, the continuous-time spectrum based models have also been proposed. Nutting[42] proposed a general relationship of stress-strain with time domain, in which the strain is expressed as a function of stress multiplied by the power function of time. The time dependent power law has been used to express the relaxation function[43] as follows:

$$E(t) = \frac{1}{1+\left(\frac{t}{\tau}\right)^{\delta}} \tag{5}$$

where $\tau$ is a characteristic time in second.

The power law model has the advantage of simplicity by using only a few model parameters to describe the relaxation modulus. However, it is unable to accurately capture the modulus at the high or low time ranges[44].

The sigmoidal function has also been used to describe function relaxations:

$$S(t) = \frac{1}{1+e^{-t}} \tag{6}$$

Sigmoidal function describes an S-shaped curve that is somehow similar to $E(t)$ at the logarithmical time scale. Accordingly, a sigmoidal function-based mathematical model has been proposed to fit the experimental data of the absolute values of complex modulus (named as "dynamic modulus") in frequency domain[45,46] as follows:

$$\log(|E^*(f)|) = \delta + \frac{\alpha}{1+e^{\beta+\gamma\log(f)}} \tag{7}$$

where $E^*$ is complex or dynamic modulus, $\delta$, $\alpha$, $\beta$, and $\gamma$ are fitted model parameters, and $f$ is frequency.

This model has been successfully implemented in fitting on the dynamic modulus of asphalt or bitumen materials[47]. However, this model lacks physical mechanism to describe the material viscoelastic behaviors, e.g. it only captures the absolute value of complex modulus mathematically. As a result, it has not been numerically implemented for simulations of deformations of structural system.

By changing and/or extending the continuous-time-spectrum based formulas to introduce new models, it can improve the model accuracy in fitting and predicting experimental data of modulus and simulating responses of materials and structures. With the physical mechanism also considered, the developed model could describe the viscoelasticity of general materials at both the experimental and numerical scales to improve the accuracy of existing models. Based on validation results, Equation (5) and (6) are relatively simple, but unable to produce sufficient accuracy in fitting experimental data. Equation (7) is a purely mathematical function and lacks physical mechanism as discussed above. As trial results based on experimental data of different materials including infrastructure materials, polymers, tissues and biomaterials, we propose a new material viscoelastic model to simulate stress relaxation and responses (i.e. deformation) of general materials. The main body of the model formula is extended from the sigmoidal function at the logarithmical scale but considering the nonlinear strain hardening or softening as follows:

$$\log E(t) = \log E_\infty + \log(E_0 - E_\infty)(1+\varepsilon)^{\beta} \frac{1}{1+\mu e^{\alpha\log(t/\tau_0)}} \tag{8}$$

where $E_\infty$ is the minimum modulus at the rubbery stage and it can be zero dependent on materials, $E_0$ is the instantaneous modulus at glassy stage (purely elastic), $\alpha$ is a power term that determines the rate of modulus decay with time, $\mu$ is a temperature-dependent model parameter that converts the temperature effect to relaxation time, $\tau_0$ is a characteristic time in second to vanish the unit of $t$ (e.g. $\tau_0 = 1$s and $t/\tau_0$ is an unitless number), and $\beta$ is a strain hardening or softening coefficient and the model becomes linear viscoelastic when $\beta = 0$.
It shall note that in this model the exponential term ($e^{\log(t/\tau_0)}$) can be converted to a power-law term $(t/\tau_0)^{u/ln10}$ without using the logarithmical term for $t/\tau$. However, the model formula with logarithmic scale applied on both $E(t)$ and $t/\tau$ can fit experimental data more accurately than the non-log model based on validation results of different materials. In addition this model is not considering plasticity which more often has the strain hardening or softening behaviors.
To simplify the model format, it can be expressed as follows given that both modulus and time are in the logrithimical scale:

$$E(t) = E_\infty + \Delta E(1+\varepsilon)^\beta \frac{1}{1+\mu e^{\alpha \log(t/\tau_0)}} \tag{9}$$

where $\Delta E = E_0 - E_\infty$ is the modulus variation range.
To consider the temperature effect, a shift factor $\alpha_T$ is applied to convert the physical time to a reduced one $t_r$ at the reference temperature according to the temperature-time superposition principle, and thus $E(t)$ is re-expressed as:

$$E(t) = E_\infty + \Delta E \frac{1}{1+\mu e^{\alpha \log(\alpha_T t/\tau_0)}} (1+\varepsilon)^\beta$$

$$= E_\infty + \Delta E \frac{1}{1+\mu e^{\alpha \log(t_r/\tau_0)}} (1+\varepsilon)^\beta \tag{10}$$

where $t_r$ is the reduced time.
$\alpha_T$ can be expressed as a function of temperature following the Arrenius law as follows:

$$\log(\alpha_T) = -\frac{E_a}{R}\left(\frac{1}{T+273.15} - \frac{1}{T_0+273.15}\right) \tag{11}$$

where $T$ is temperature ($^o$C) and $T_0$ is reference temperature ($^o$C), $E_a$ is a fitted model parameter - an "energy" ($\mathrm{J.Mol^{-1}}$) required for thermal transition, $R$ is a constant (like the gas constant, $\mathrm{J.K^{-1}.Mol^{-1}}$), and $E_a/R$ is a model parameter index (K).
Other formulas for calculating $\alpha_T$ have also been proposed including the WLF (Williams-Landel-Ferry) rule[48] and the polynomial function[46]. The formula of $\alpha_T$ shall be properly chosen for the specific material in order to achieve satisfied accuracy in fitting experimental data.
Though the dynamics of polymer networks is complex and not completely understood in polymer physics, micro-mechanisms and network-based models have been proposed for simulating material viscoelasticity primarily for polymers, rubber materials and elastomers. Model examples include the coarse-grain polymer network model in which the mesh-like networks are stretched in the viscous medium[49]; the symmetrically growing treelike micro-network model using sub-chains and the non-affine microspheres[50]. The network may be modified such as including the crosslinked kinetics into polymers[7].

However, for general materials other than polymers/rubbers such as the amorphous asphalt concrete, the multi-molecular-chain-based models could not properly explain the physical mechanism. In order to interpret the physical mechanism for general materials, we propose a network-based model with a relatively simple format to model a broad range of general materials including the macromolecular based polymers and bio-materials, amorphous and crystal or polycrystalline materials, and composites as shown in **Figure 2**. Examples include the agar made from algae –a seaweeds, polymer network, spider silk - a protein fiber consisting of macromolecular, organic asphalt, and bone tissue – a composite material.

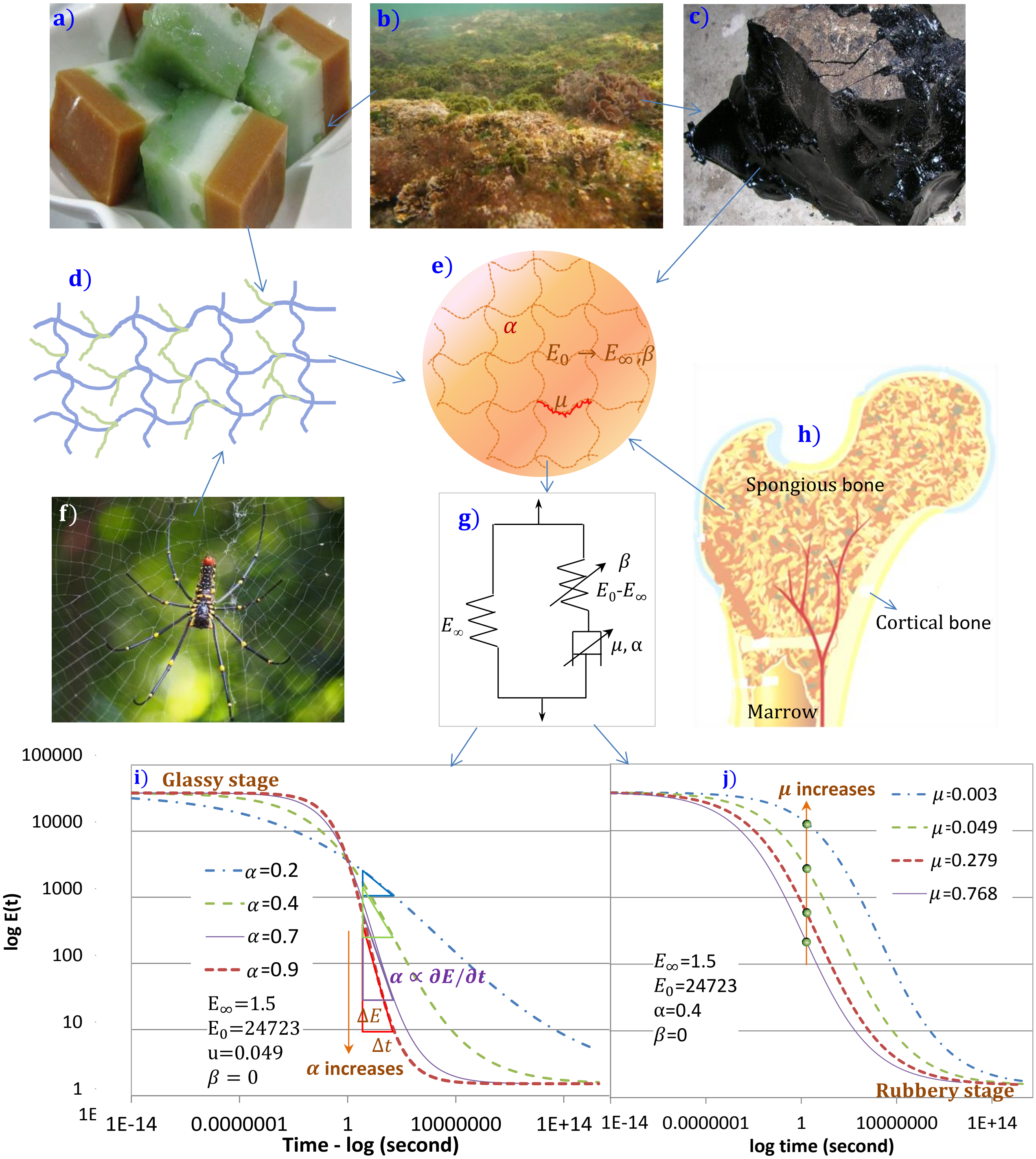

**Figure 2.** Viscoelastic materials and proposed general model: a) agar cake made from algae; b) algae on sea bed (courtesy of Toby Hudson); c) natural asphalt formed from the remains of ancient microscopic algae; d) polymer network with linear, cross-linked and branched chains; e) proposed network-viscous medium model; f) photo of a robust spider silk web; g) a two-element model system; h) effects of model parameter $\boldsymbol{\alpha}$ and i) effect of model parameter $\boldsymbol{u}$ on relaxation modulus $\boldsymbol{E(t)}$.

The proposed model can be physically presented by an elastic spring network with viscous medium filled between the network meshes as illustrated in Figure 2e. The model formula can be mathematically derived from a two element system (variable dashpot and spring) as shown in Figure 2g with proof shown in Appendix A. The elastic network carries a modulus capability $E \in [E_\infty, E_0]$ which transits between the glassy and rubber stage, and it may display strain hardening or softening when strain becomes large as characterized by the $\beta$ coefficient. According to the temperature-time ($T - t$) superposition rule, the temperature effect is equivalent to the reduced relaxation time (i.e. a higher $T$ corresponds to a higher $t$ and vice versa). With the increase of temperature or relaxation time, the model domain becomes "softer" and the value of $E(T)$ decreases. At the time of zero or infinitely low temperature of glassy stage, the network carries the highest modulus of $E_0$, and with the increase of relaxation time or temperature the network modulus decreases continuously until reaching $E_\infty$ at the infinite time of rubbery stage (see Figure 2i). The model parameters also include $\mu$- a coefficient of friction between the elastic network and viscous medium and $\alpha$ - a parameter indicating temperature dependency or thermal sensitivity of the viscous medium (see Figure 2e). $\alpha$ controls the slope or rate of the modulus change such that $\alpha \propto \partial E(t)/\partial t$. A higher $\alpha$ value indicates a higher thermal sensitivity that $E(t)$ will relaxes or reduces more rapidly within the same time or temperature difference as shown in Figure 2i. A higher $\mu$ value indicates a higher friction force with increased contribution of the viscous medium to the system response, resulting in relatively higher viscosity of the material system. With the increase of $\mu$, $E(t)$ increases due to the increased "friction force" as shown in Figure 2j.

Time or temperature or strain rate dependency of viscosity has been considered such including the Non-Newton fluids. The time or temperature dependency of viscosity could be expressed using mathematical formula such as those following the power-law function of strain rate[51] or linear function of time[52]. Other studies including Lion's[53] rheological model have also considered the time and temperature dependent viscosity. By using the time and temperature dependent viscosity, the mathematical formula of the model can be proven based on the two-element system shown in Figure 2g with details presented in Appendix A.

To ensure the thermodynamic consistency, the Calusius-Duhem inequality shall be satisfied[54] as proven in Appendix B. In the three-dimensional (3-D) domain, the shear and bulk relaxation moduli ($G(t)$ and $K(t)$) are required inputs for simulating responses of materials. With the thermodynamic consistency, the shear and bulk relaxation moduli can be expressed in the same formula as that of $E(t)$:

$$G(t) = G_\infty + (G_0 - G_\infty)\frac{1}{1+\mu_G e^{\alpha_G \log(t/\tau_0)}}(1+\varepsilon)^{\beta_G} \tag{12}$$

$$K(t) = K_\infty + (K_0 - K_\infty)\frac{1}{1+\mu_K e^{\alpha_K \log(t/\tau_0)}}(1+\varepsilon)^{\beta_K} \tag{13}$$

where $G_\infty$ is shear modulus at infinite time ($t = \infty$), and $G_0$ is the instantaneous shear modulus, $K_\infty$ is bulk modulus at infinite time ($t = \infty$), $K_0$ is the instantaneous bulk modulus, and $\alpha_G$, $\alpha_K$, $\beta_G$, $\beta_K$, $\mu_G$ and $\mu_K$ are model parameters.

## III. EXPERIMENTAL VALIDATION

We validate the new model with experimental data of different materials, including the infrastructure material, polymer, biomaterial, and tissue. The experimental validations include both the model fitting and predictions. When part of the experimental data within a middle time range (training data) is used for fitting model parameters, the rest of the data at the lower and higher time ranges (testing data) is predicted using the fitted model parameters.

### A. Improves stability and convergence in fitting laboratory data $E(t)$

One of the most popular optimization methods, the nonlinear reduced gradient method was used to fit the model parameters by minimizing the objective function:

$$\min f(x_1, x_2, \ldots, x_m) \ \forall\, x_i > 0 \tag{14}$$

$$f(x_1, x_2, \ldots, x_m) = \sum_i^n \left[E_i(x_1, x_2, \ldots, x_m) - \hat{E}_i\right]^2 \tag{15}$$

where $f$ is the objective function, and $x_i$ is the model parameters (e.g., for PS model$x_i = E_i$ and $\eta_i$), $n$ is the number of experimental data points, $E_i$ and $\widehat{E_i}$ is the $i^{th}$ modeled and measured modulus, respectively.

For this optimization method, the gradients (derivatives of $f$ with respect to each model parameter) are calculated using the central finite difference (CFD) method which plays a crucial role for the iteration results. The iteration goal is to satisfy the first order essential optimal condition (i.e. the gradient equals or is close to zero). Here the reduced Chi-square is used to evaluate the goodness of fit for the objective function as follows:

$$\chi^2 = \frac{\chi^2}{n} = \frac{1}{n} \sum_i \frac{f(x_1, x_2, \ldots x_m)}{\mathrm{Var}} \tag{16}$$

wehre Var is the variance of measurements.

A lower $\chi^2$ value indicates a higher fitting accuracy. It is known that inverse computation is generally dependent on seed values and it may turn out multiple results of model parameters which all satisfy the first ordr optimal condition[21], which is especially true for the model with relatively large number of model parameters such as the PS model studied here. All modeling results are compared with the PS model, and to be "fair" for only consdiering linear viscoelasticty in the proposed model $\beta$ value is considered zero for all the tests comparing PS as discussed later. **Figure 3** illustrates the optimization results for PS with $n = 2$ and the proposed model (with the same number of model parameters) using diffeent group of seed values (the initial trial values) consdiering three general cases: 1) fitted modulus valus are higher than, 2) lower than, and 3) close to measurment values usigng the seeds. Results have shown that PS model has produced varaible results (i.e. different $E(t)$ shapes) when using variable seed values (see Figure 3a). However, the proposed model yilds almost unique solution for these three seed groups, indicating its higher stability and convergence (Figure 3b). When using a relatively large term number $n$ for PS,its variability can be greater dependent on the $\tau_i$ values since it could be more difficult to estimate proper $\tau_i$ seed values for achiving unique solutions. In comparison, for the porposed model it is more easy to achieve unique solution as well as to estimate proper seed values. For example, its $E_\infty$ and $E_0$ values can be properly estimated according to the value range of exopeimrental data (i.e. the minimum and maximum value is relatively close to $E_\infty$ and $E_0$, respectively). Therfore, the proposed model achieves more stable and unique, yet accurate solutions.

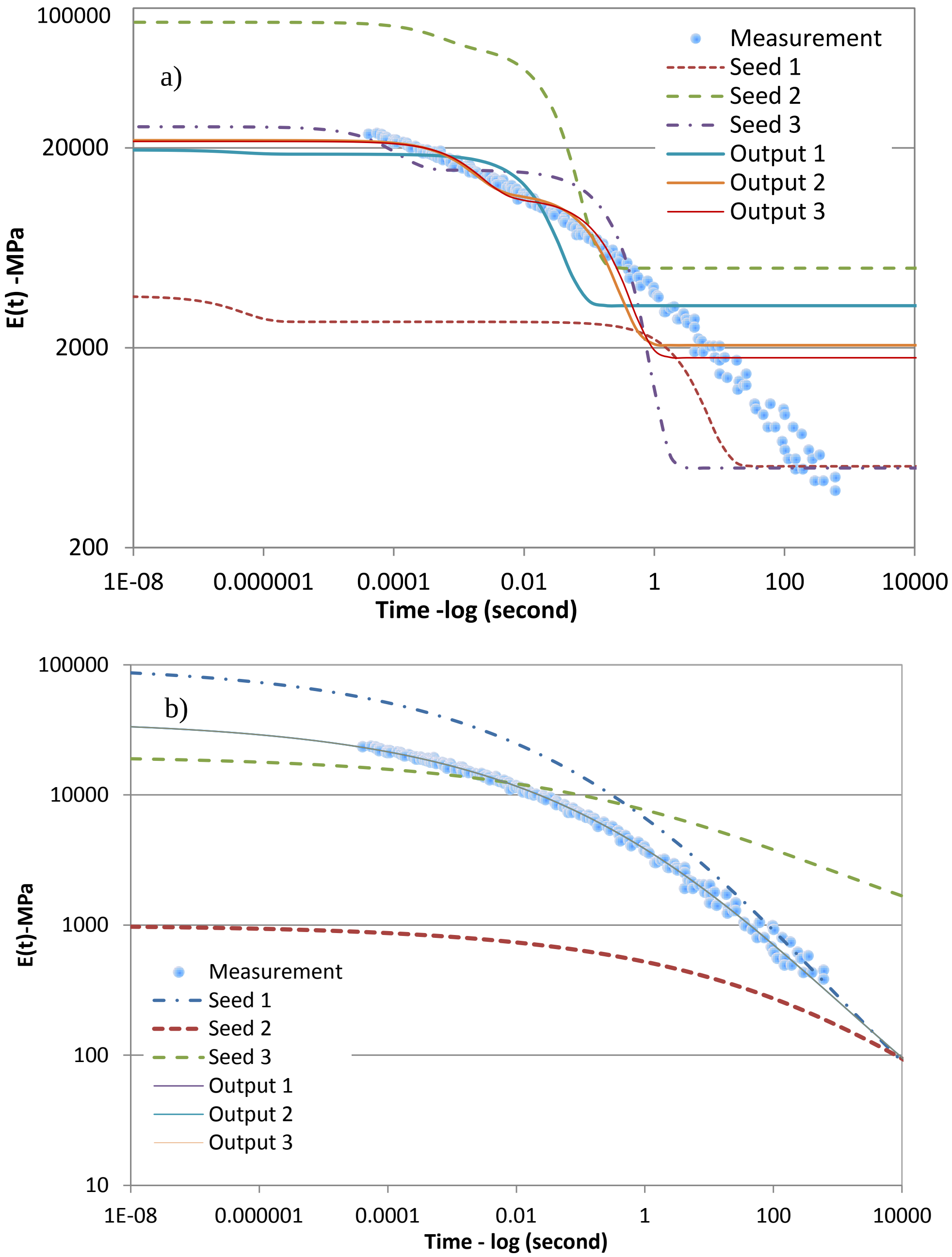

**Figure 3**. Optimization analysis to fit experimental data: a) Prony series and b) proposed model both with three seed inputs and the proposed one shows more robust and accurate convergence than PS model.

### B. Improves accuracy in fitting and predicting $E(t)$ outside of experimental range

We performed experimental tests on the relaxation modulus of an AC mixture, one of the mostly used infrastructure materials. AC can be regarded as a three-phase composite, consisting of the viscoelastic asphalt binder or asphalt mastic (asphalt with sands and minerals), elastic aggregates, and air voids (see Figure 4). Asphalt is a heavy organics presented in petroleum and performs like a viscous fluid at high temperatures while a viscoelastic solid at normal or low temperatures. Asphalt is amorphous with low molecular weight but very complex or variable molecular structures, and it cannot be separated into individual components or narrow fractions[55] (see the molecular structure of one crude asphalt material in Figure 4). Therefore, its viscoelastic behaviors may not be properly interpreted by the molecular chain segments of polymer model as discussed early. This is also true for many other materials with different morphology than polymers and rubber-like materials.

In **Figure 4** we evaluate the fitting accuracy of PS model as compared to the proposed model using lab measurements based on the AC material. Experimental measured $E(t)$ of AC specimen are fitted by PS with a variable number of $n$ ranged from 1 to 30 (total 3 to 61 model parameters). Results have shown that with the increase of $n$, the model intends to capture a wider range of time and modulus values with improved fitting accuracy. This is the accumulation results of modulus values distributed at variable relaxation time spectrums. Results have also shown that when $n$ is relatively large (i.e. $n \geq 14$) the $\chi^2$ value almost has no change any more without improving fitting accuracy in response to $n$ value. However, different seed values result in different model shapes although with very close $\chi^2$ values (the same or similar fitting accuracy). For example, when using relatively large seed values of $\sum E_i$ the predicted modulus values at the low time range outside of the experimental range are larger than those predicted using smaller seed values of $\sum E_i$, but both have the same or very close $\chi^2$ values (see Figure 4: $E(t)$ with $n = 17^a$ *vs.* those with $n = 14$ and $n =$30).

For PS model "oscillations" (unsmooth curves) are observed due to its finite terms in discrete spectrums of relaxation time. With the increase of $n$, the curve of $E(t)$ can become smoother when using proper seed values, but it also becomes more difficult to estimate plenty of seed values, as well as different seed inputs may turn out different $E(t)$ shapes (no "unique" solutions). In addition, some sharp transitions to the glassy and rubbery stages are observed once the model reaches beyond the experimental range (see Figure 4). In comparison, the proposed model can achieve a smoother but accurate curve fitting, and it intends to transit to the glassy and rubbery stages smoothly by capturing the full range of relaxation time. It shows that the propose model may result in higher $E(t)$ values than that of PS at the low time range dependent on the seed values of PS, and further evaluation will be conducted through model predictions in the next.

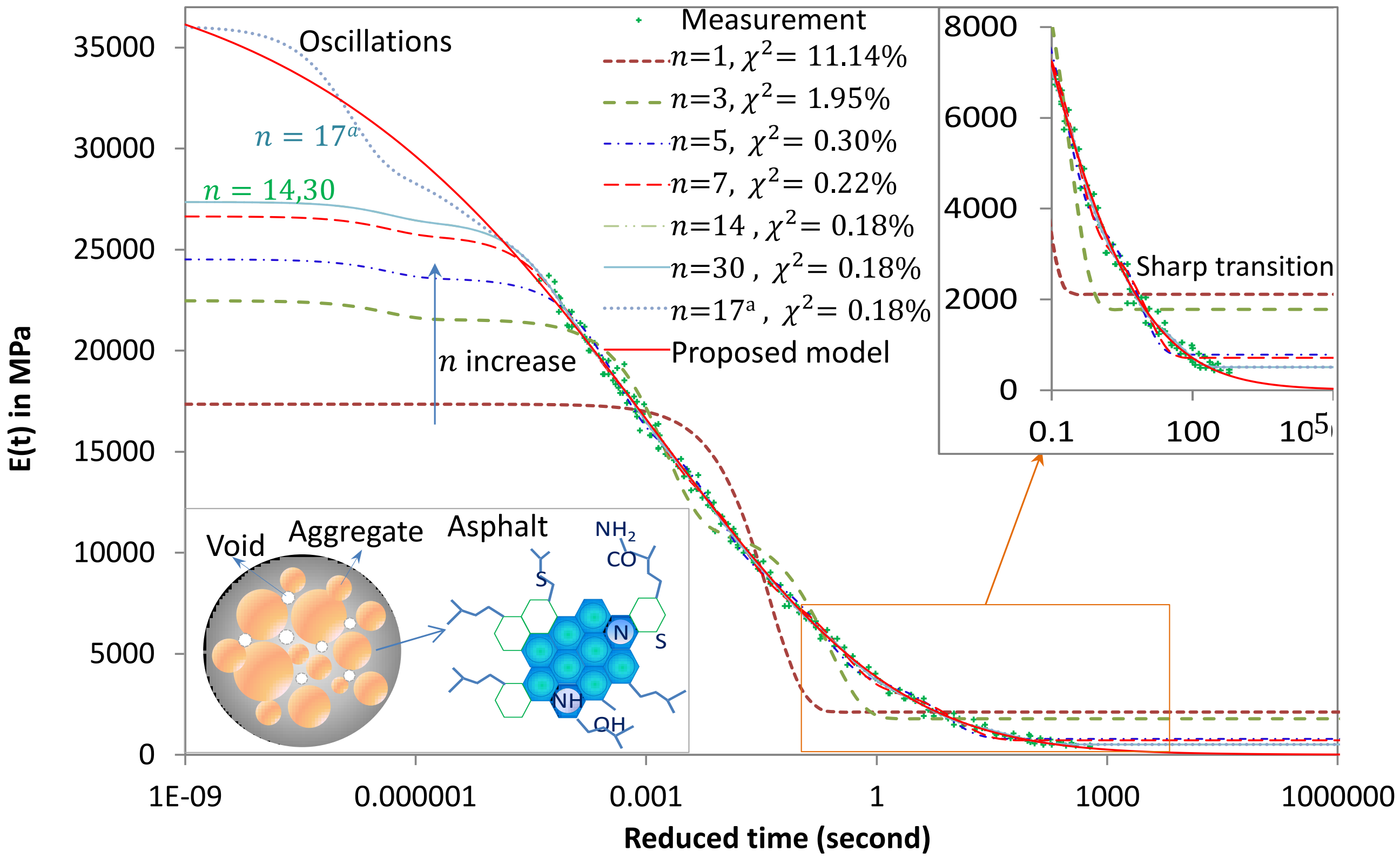

Figure 4. Laboratory test results and model fit of asphalt concrete.
[a] using larger seed values of $\sum \boldsymbol{E_i}$ for the curve with $\boldsymbol{n}$=17.

In the following we present important tests to validate the new model on another three materials: shape-memory polymer, agar and bone. For PS model, a term number of $n = 14$ is used for all materials in the following sections with satisfied accuracy. When only part of the experimental data at the middle range of relaxation time are used as training data for fitting model parameters, the rest are predicted using the fitted model parameters.

**Figure 5** presents the model fitting and prediction of the shape-memory polyurethane (PU) polymer. PU material is composed of chains of organic units joined by the urethane links 54 (see Figure 5a). With the deformation memory effect it can be used for making foam memory mattress which adapts to the shape and weight distribution of human body for comfortable sleep. $E(t)$ of PU was measured at six temperatures of -30, -22.5, -20, -17.5, -15, and -12.5°C 58. Results show that the proposed model can accurately fit the experimental data which are "shifted" to that at the reference temperature of -17.5°C following the proposed model formula (see Figure 5a). In comparison, for PS model a "sudden" oscillation appears at the low time range $t \in (10^{-5}, 10^{-3}$ seconds) outside of the experimental range (see Figure 5b). However, PS still achieves very satisfied fitting accuracy with a $\chi^2$ value of only 0.65%. This further indicates its lower stability and higher dependency on seed values especially for the modeling range outside of the experimental data. The proposed model slightly under-predicts $E(t)$ at the high time range, but has significantly improved the prediction accuracy as compared to that of PS. PS intended to more rapidly converge to the glassy and rubbery stages once it reaches out of the experimental range, and thus produced prediction errors at the low and high time ranges (see Figure 5b).

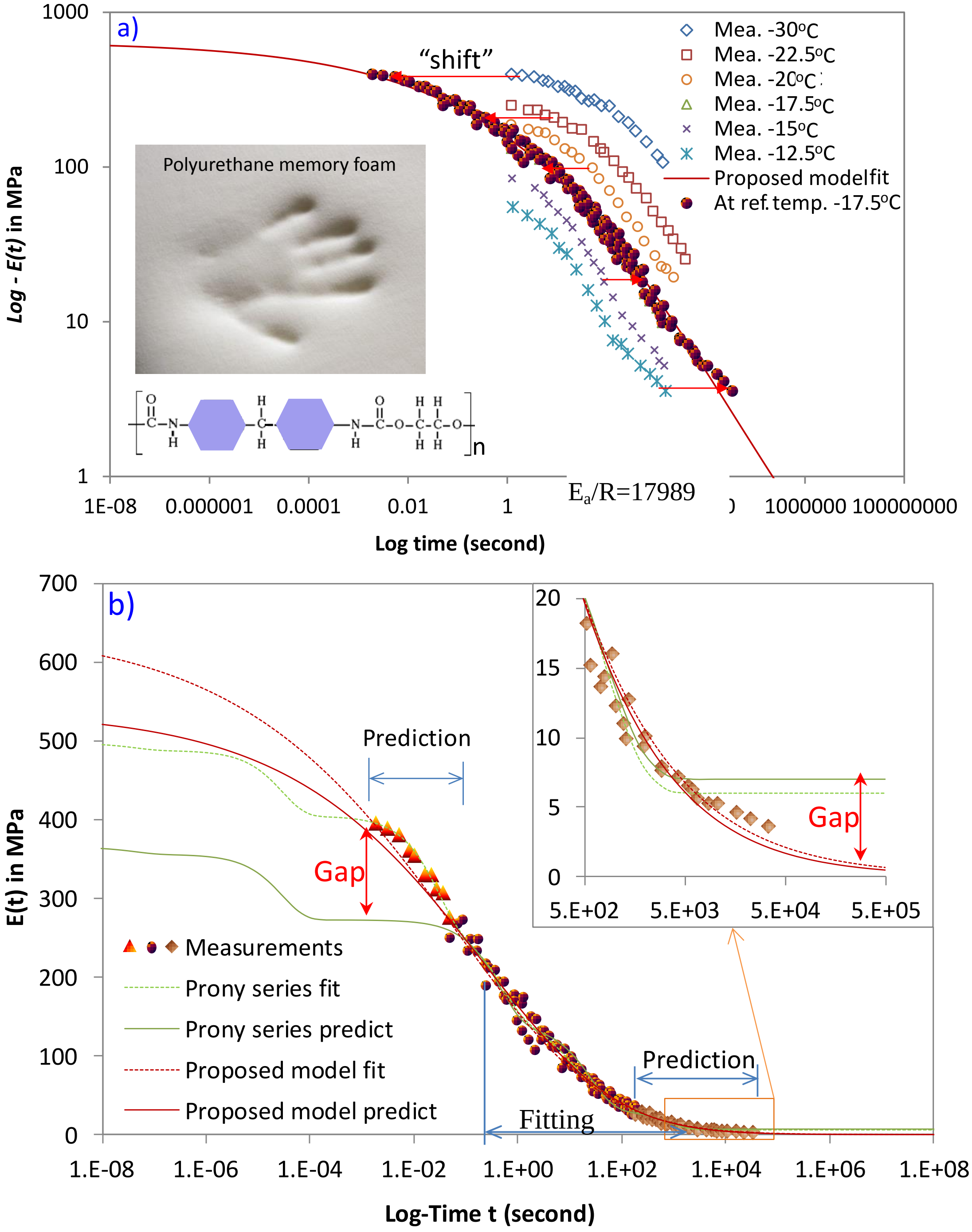

**Figure 5**. Shape memory polymer measured relaxation modulus and modeling results: a) lab test results (reproduced with permission from 58) and data fitting using proposed model through "shifting" data at different temperatures to that at the reference temperature according to the temperature-time superposition rule and b) model fitting and predictions using both the proposed model and GM model and the proposed one demonstrated higher accuracy.

**Figure 6** presents the model fitting and prediction for the bio-/tissue materials of agar and bovine femoral bones. As a biomaterial derived from agarose of raw seaweed, agar can be used to make foods such as agar cake. Figure 6a shows that the proposed model and PS for argar materials have produced relatively close results on both fitting and predictions, which is due to the relatively high linearity of experimental data and less variability of modeling. PS model has slightly overpredicted $E(t)$ at the high time range.

Bone, a rigid and relatively light organ, has been often considered as a composite material primarily consisting of hydroxyapatite-like mineral particles (e.g. calcium) embedded in a plaint matrix of collagen fibers[27]. The cortical or compact bone- the dense outside layer (see Figure 2h) has noticeable viscoelasticity as contributed by the collagen fibres and non-fibrous proteins in the bone matrix[8]. As shown in Figure 6b1, the PS model using relatively high $\sum E_i$ seed values produced some curve "oscillations" to fit experimental data of a cortical bone, but it has fairly satisfied and very close fitting accuracy as that using lower $\sum E_i$ seed values (see Figure 6b2), e.g $\chi^2$ value of 0.34% and 0.21%, respectively. In comparison, the proposed model produces a smoother and more accurate curve fitting, and it estimates $E_0$ at zero time as16.5 GPa which falls within the range measured by previous researchers, i.e. 11-21 GPa[57]. Results have also shown that the proposed model has significantly improved the prediction accuracy. For example, PS over-predicts $E(t)$ at the high time range and produced obvious gaps as it intends to more rapidly converge to $E_\infty$ outside of the fitting range (see Figure 6b2).

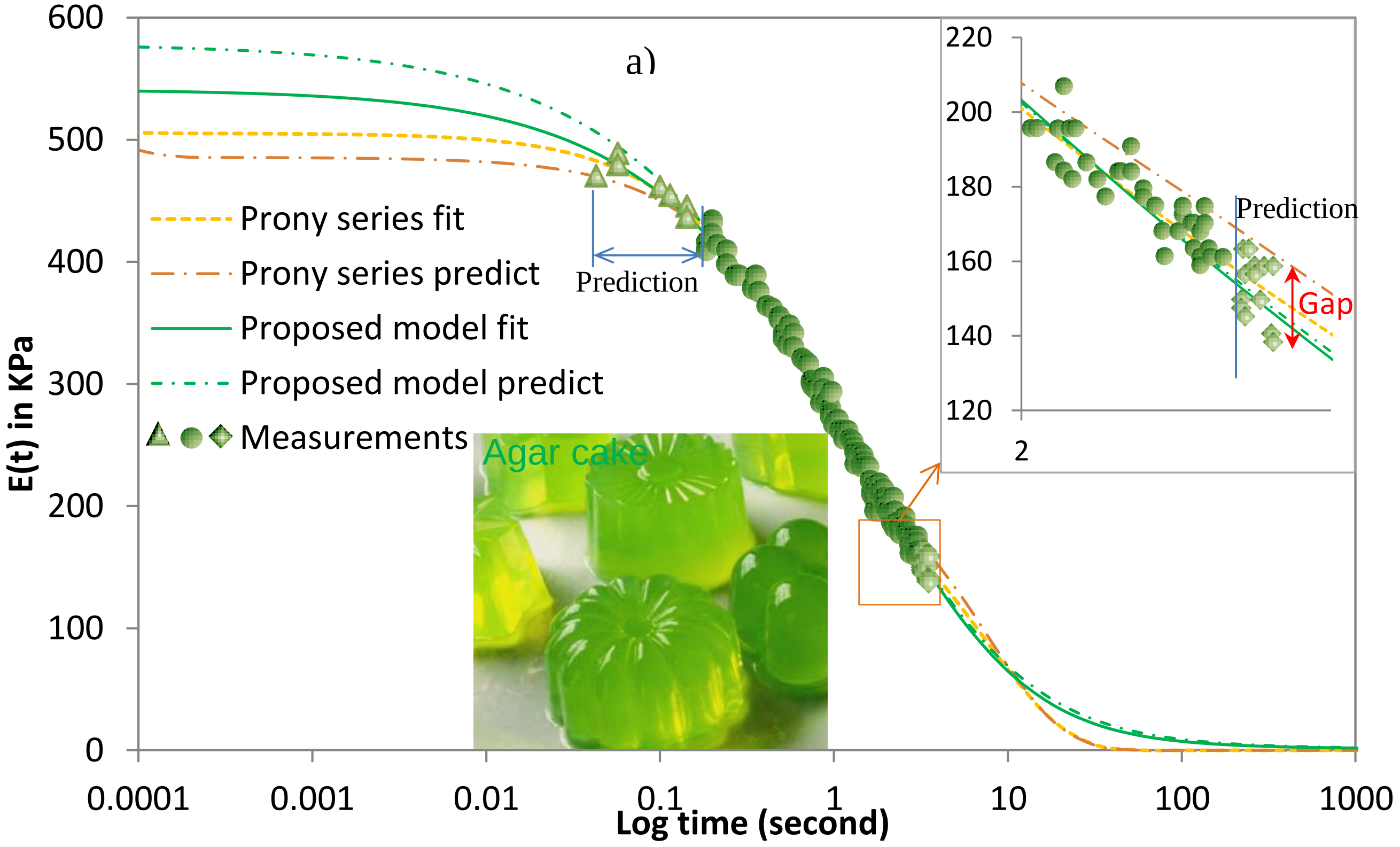

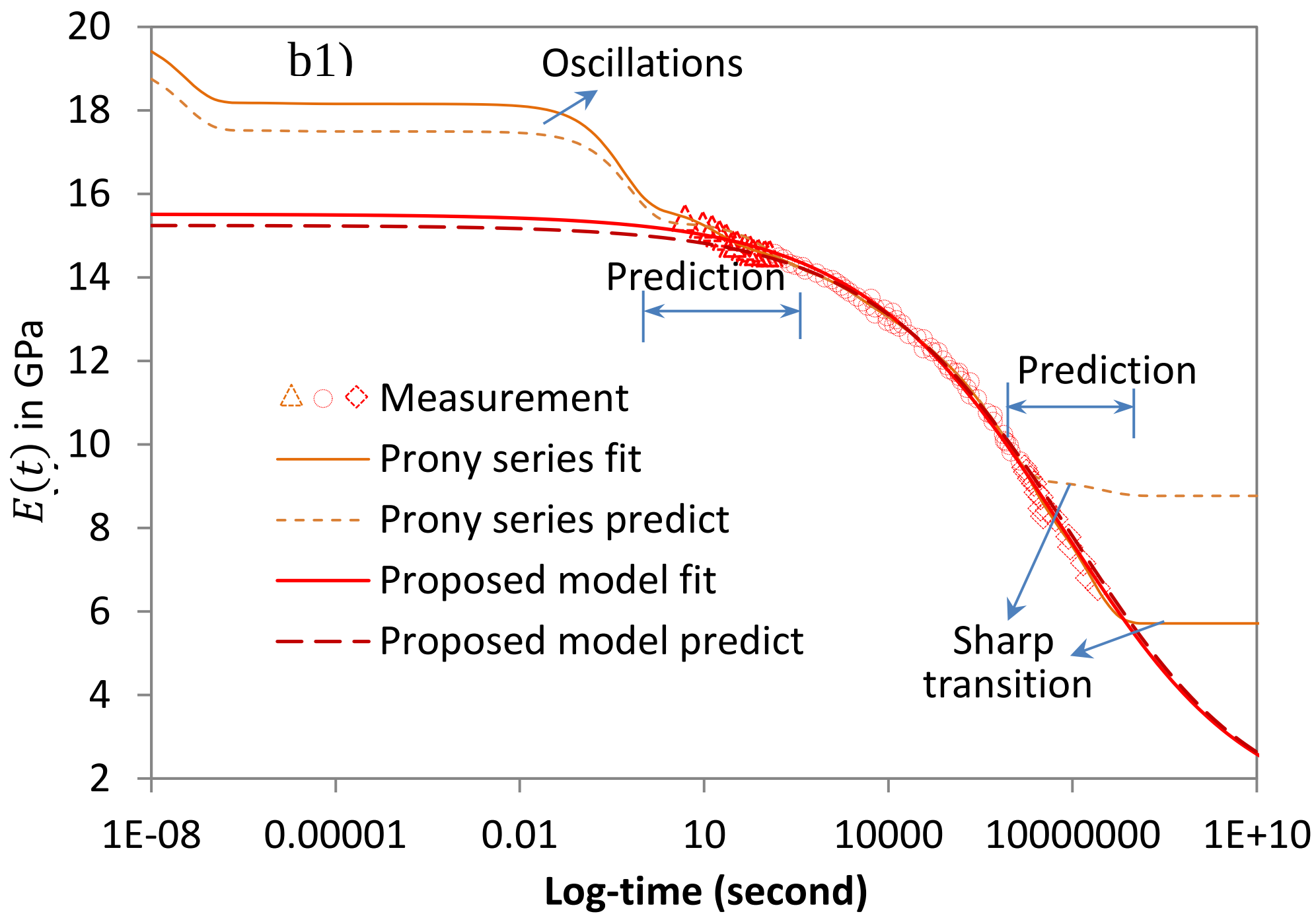

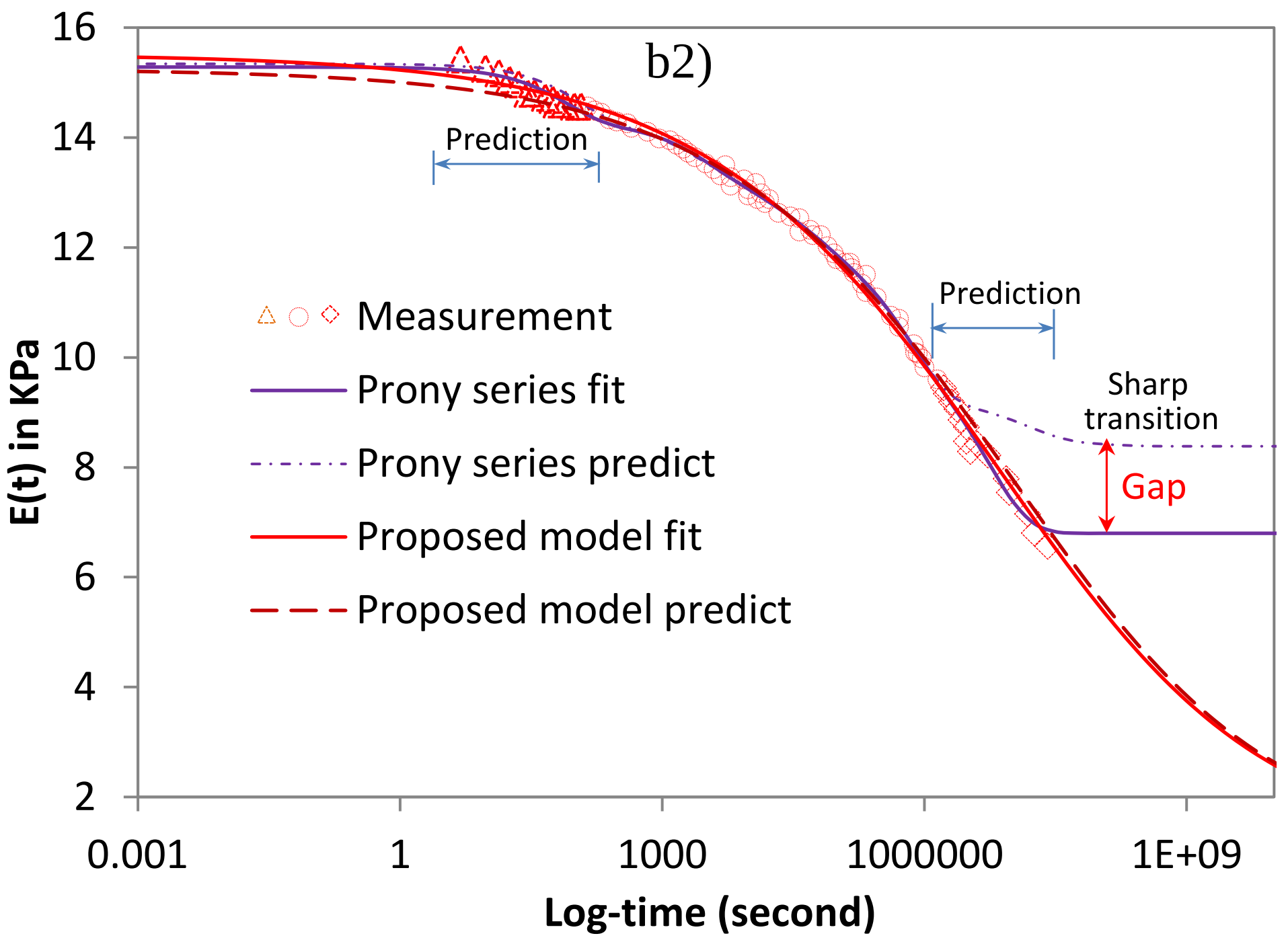

**Figure 6.** Data fitting and prediction for bio-/tissue materials: a) model fit and prediction of agar material (measurement data reproduced with permission from 59) and b) model fit and prediction of bovine femoral cortical bones: (b1) using higher $\sum \boldsymbol{E_i}$ seed values and (b2) using lower $\sum \boldsymbol{E_i}$ seed values, both materials show showing the proposed model yields higher accuracy than PS in both fitting and prediction outside of experimental measurement range (measurement data reproduced with permission from 8).

The root mean square (RMS) can be used to evaluate the prediction accuracy as follows:

$$\mathrm{RMS} = \sqrt{\sum\left[\left(E_i - \hat{E}_i\right)/E_i\right]^2/N} \quad (17)$$

where $E_i$ is the predicted modulus and $\hat{E}_i$ is the measured ones, and $N$ is the number of data points.

**Table 1** lists the fitted model parameters and RMS values of prediction for materials studied in this research. The bone is the stiffest material with the highest $E_0$ and $E_\infty$ values and poses the longest relaxation time spectrum (see Table 1a), while agar is the softest one with the lowest modulus value. Agar poses a higher $\alpha$ value than others, indicating its relatively higher viscosity-dependent temperature sensitivity, followed by the PU and then AC, and lastly the bone material which is most elastic or least viscous relatively. Table 1b presents the RMS values of predictions (calculated based on the data points of predictions only). Results show that the proposed model has improved prediction accuracy as compared to PS. RMS values of the proposed model are fairly small for the Agar and Bone materials (4.84% and 2.90%, respectively), indicating very accurate predictions.

**Table 1**. a) Model fitting parameters of proposed model

| Material | AC | PU | Agar | Bone |
|---|---|---|---|---|
| $E_\infty$ (MPa) | 6.50E-01 | 8.68E-03 | 1.02E-03 | 1.00E+03 |
| $E_0$ (MPa) | 6.12E+03 | 5.11E+02 | 5.30E-01 | 1.65E+04 |
| α | 0.3800 | 0.4186 | 1.4517 | 0.5619 |
| $\mu$ | 8.19E-01 | 5.40E-04 | 5.65E-01 | 7.73E-01 |

b) RMS of predictions

| **Model\Material** | **Agar** | **PU** | **Bone** |
|---|---|---|---|
| Prony series | 8.55% | 68.08% | 13.09% |
| Proposed model | 4.84% | 14.45% | 2.90% |

Note: for the proposed model $\beta = 0$ in order to compare with the PS within lienar viscoelasticity

One may argue that given a very large term number for the PS model (e.g. $n > 30$) the modeling accuracy could be improved. There are also existing methods developed to improve the accuracy of PS fitting by using the pre-smooth method[39]. However, the proposed model could be superior than the PS regarding the following aspects: 1) a model with less parameters yet satisfied accuracy is always preferred for simplicity; 2) a large number of model parameters has lower stability and higher variability for data fitting and prediction; 3) it becomes more difficult to explain the physical mechanism when model parameters become large. For example, the spring-dashpot systems of the PS may interpret the multi-chains of polymer and other macromolecular materials, but not for general materials with different morphology and 4) PS model is not accurate for prediction outside of experimental range, which was also proved by previous study[3239], while the proposed model has improved the prediction accuracy. The prediction accuracy could be very important since laboratory testing has limitations or difficulty in measuring modulus at the low and high frequency or time ranges, which are usually determined from model predictions and are often required inputs for modeling responses, evaluation and design of materials.

## IV. NUMERICAL SOLUTION METHOD

We develop a Galerkin finite element (FE) method to implement the proposed model and solve the resulting partial differential equations for simulating material responses under both static and dynamic loading as derived in Appendix C. We have also conducted error analysis to compare the numerical accuracy of the proposed model and that of PS.

### A. A Galerkin-based numerical solution method within time-domain

For solids in the continuum state, the strong form of the governing state equation for the dynamic problem can be formed as follows:

$$\boldsymbol{\nabla} \cdot \boldsymbol{\sigma} + b = \rho\, \ddot{\boldsymbol{u}}(t) \quad \text{on } \Omega \in \mathbb{R}^3 \times [t_0, t_d] \tag{18}$$

where $\boldsymbol{\sigma}$ is stress tensor, $\boldsymbol{u}$ is displacement, $b$ is body force, $\rho$ is material density, $t$ is time variable, $\Omega \in \mathbb{R}^3$ is a 3-D space domain, and $[t_0, t_d]$ is a time domain.

The viscoelastic stiffness matrix implementing the new material model is created (see Appendix C). One shall note that numerical solution methods with PS model has been developed in existing literature[58]. PS is numerically incursive due to its exponential term for accurate integration, while the proposed model formula involves more complexity for time integration. The objective of the numerical solution with the proposed model is to achieve a compatible numerical accuracy, speed, and stability as compared to that of PS. Accordingly, we have proposed robust numerical algorithms for fast and accurate computations, and report its numerical accuracy, stability, and convergence as compared to that of PS. We proposed a combined time discretization method (Houbolt, forward and backward finite differences) to reduce time step length, and utilized the trapezoidal rule to discretize the time integration.

### B. Error analysis

The major numerical algorithm difference between the proposed model and PS lies in the time integration of the viscoelastic stiffness matrix $\boldsymbol{J}(j)$ which results directly from the mathematical formula of the model. For the proposed model the calculation of $J(j)$ time integration is based on the trapezoidal rule for discretization (see Appendix D for details), which may induce truncation errors dependent on the time step length. Using the trapezoidal rule[64] for the proposed model, the truncation error for approximating $\int_{t_1}^{t_2} J(x)dt$ within a time domain $t \in [t_1, t_2]$ divided into $N$ time steps can be calculated as follows:

$$\text{Err} = \frac{(t_2 - t_1)^3}{12N^2} J''(\xi) \quad \forall \xi \in [t_1, t_2] \tag{19}$$

where $N$ is the number of time grids.

For the PS model $J(j)$ does not produce this type of numerical error due to its recursive integration of $E(\ t)$ in the exponential format. However, PS produces computational errors due to its less accurate data fitting and predictions. Its numerical error can be calculated as follows:

$$\mathrm{Err}_{\mathrm{PS}}(t) = \mathbf{C_e} \int_{t_{j-1}}^{t_j} \left( E_{\mathrm{ps}}(t-\tau) - E(t-\tau) \right) d\tau \qquad (20)$$

where $E_{\mathrm{ps}}$ is the PS fitted modulus and $E$ is the true values, $\mathbf{C_e}$ is an elastic tensor.

The mathematical derivation for error analysis is presented in Appendix D with more details. Here we present the error analysis results of agar material as a numerical example since the $E(t)$ values of agar fitted by PS are close to that of the proposed model (see Figure 6). **Figure 7** plots the $J(t)$ errors of the proposed model versus that of PS during 10 seconds of relaxation time. Results indicate that the $J(t)$ numerical errors of the proposed model using trapezoidal rule are minimal (i.e. average 0.016% and max 0.46%). In comparison, PS model produces larger numerical errors due to its less accurate fitting on experimental data. For example, the maximum numerical error is 0.046 and 0.313 kPa.s for the proposed model and PS, respectively. The derivative of the error curve of PS model is inconsistent due to its "oscillations" of fitted $E(t)$ value as discussed early for experimental validation.

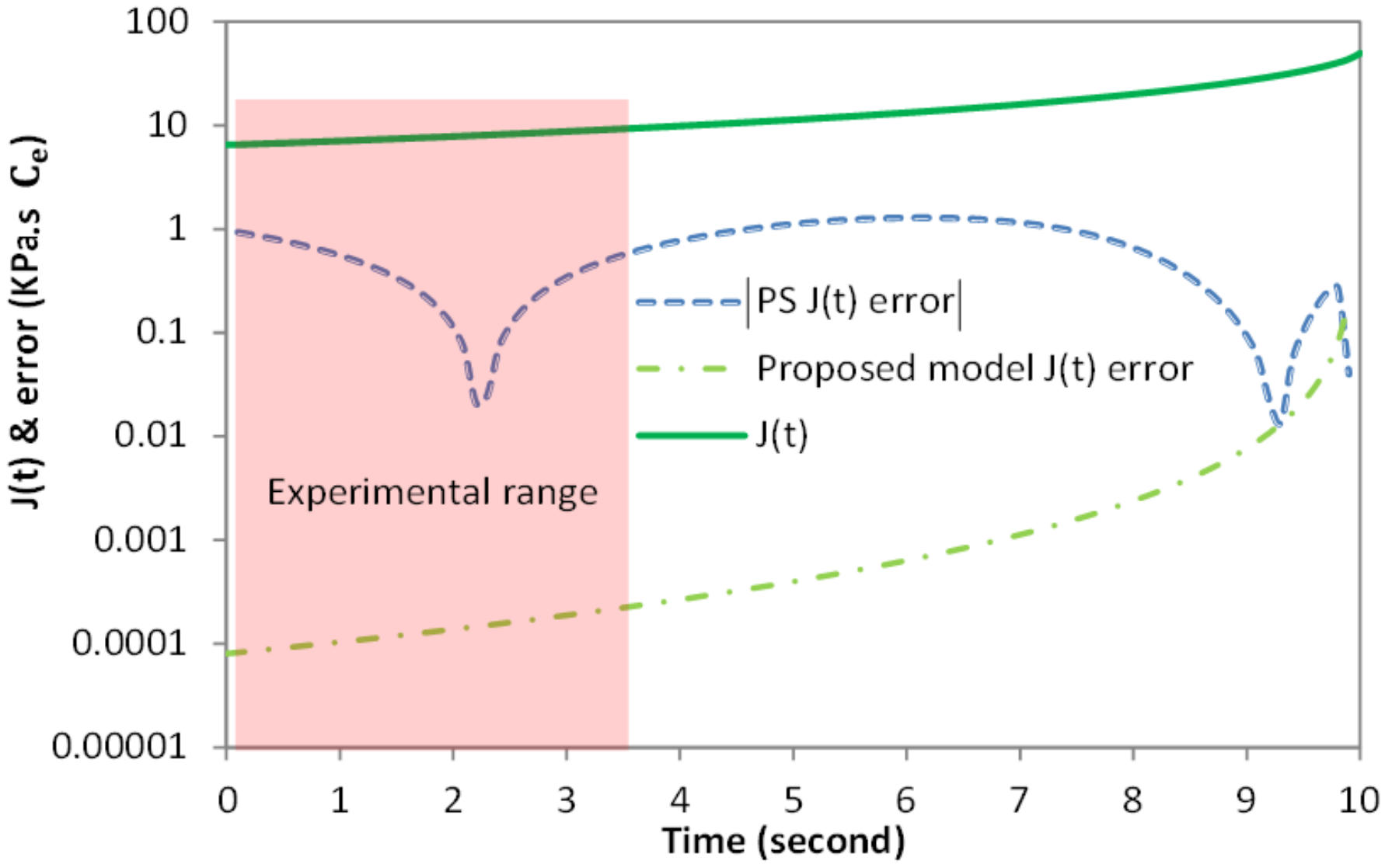

**Figure 7**. Errors of viscoelastic stiffness matrix of agar material.

## V. NUMERICAL ANALYSIS AND IMPLEMENTATION

In order to validate the accuracy of the developed FE algorithm, we first compared the simulation results of responses with that of ANSYS – one of the most powerful FE commercial software both using the PS model. They attain almost identical results although using different algorithms (e.g. different time discretization methods) with one example illustrated in section 5.2. Consequently, we implemented the proposed model in the FE algorithm for numerical analysis and compared its simulation results with that of the PS model. Lastly we implemented the model for simulating the nonlinear stress-strain behaviors of materials.

### A. Stable model with fast convergence

In **Figure 8** we report the numerical stability and convergence of the new model as compared to that of PS based on the numerical simulation results for a unit body of agar material as an illustration example. **Figure 8**a presents the simulated deformations under a constant Heaviside loading (1 kPa) for time $t \in [0,10s]$. Different time step lengths of $dt \in [0.01, 1\ s]$ are used for sensitivity analysis. Note that the initial deformation at zero time is the instantaneous elastic responses. Simulation results of both models converge at the same or very close time step length (i.e. $dt = 0.01$ s for the constant loading), although PS is slightly less sensitive to time step length. **Figure 8**b presents simulated deformations under the sinusoidal loading ($f(t) = 0.5\sin(\pi t - \pi/2) + 0.5$) with an amplitude of 1 kPa. Simulation results of both models converge at the time step length of $dt = 0.1$ s. Simulation results have illustrated that the developed model with thermodynamics consistency leads to the realistic simulations of creep and sinusoidal responses (nonnegative energy dissipation and visual work). **Figure 8**c presents the simulated deformations of the proposed model along with that of PS at a time step length of $dt = 0.01$ s. The maximum deformation differences between the proposed model and PS are -4.2% and -8.1% for the constant Heaviside loading and sinusoidal loading, respectively. This indicates that the deformation difference can be prominent at some time points, which shall cause attentions when suing PS for numerical simulation especially when the number of term is relatively small as often being the case in existing literatures (e.g. $n \leq 7$).

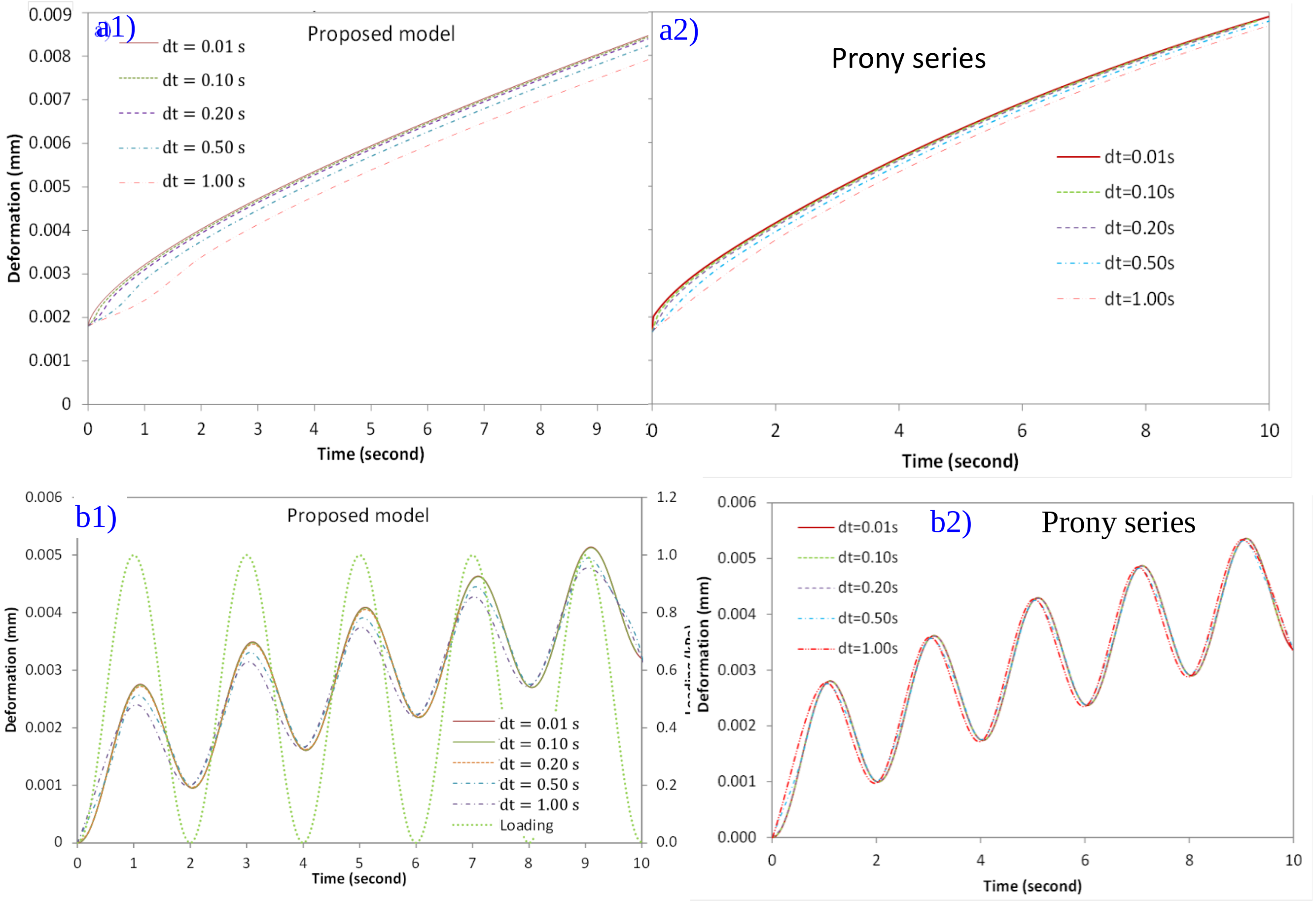

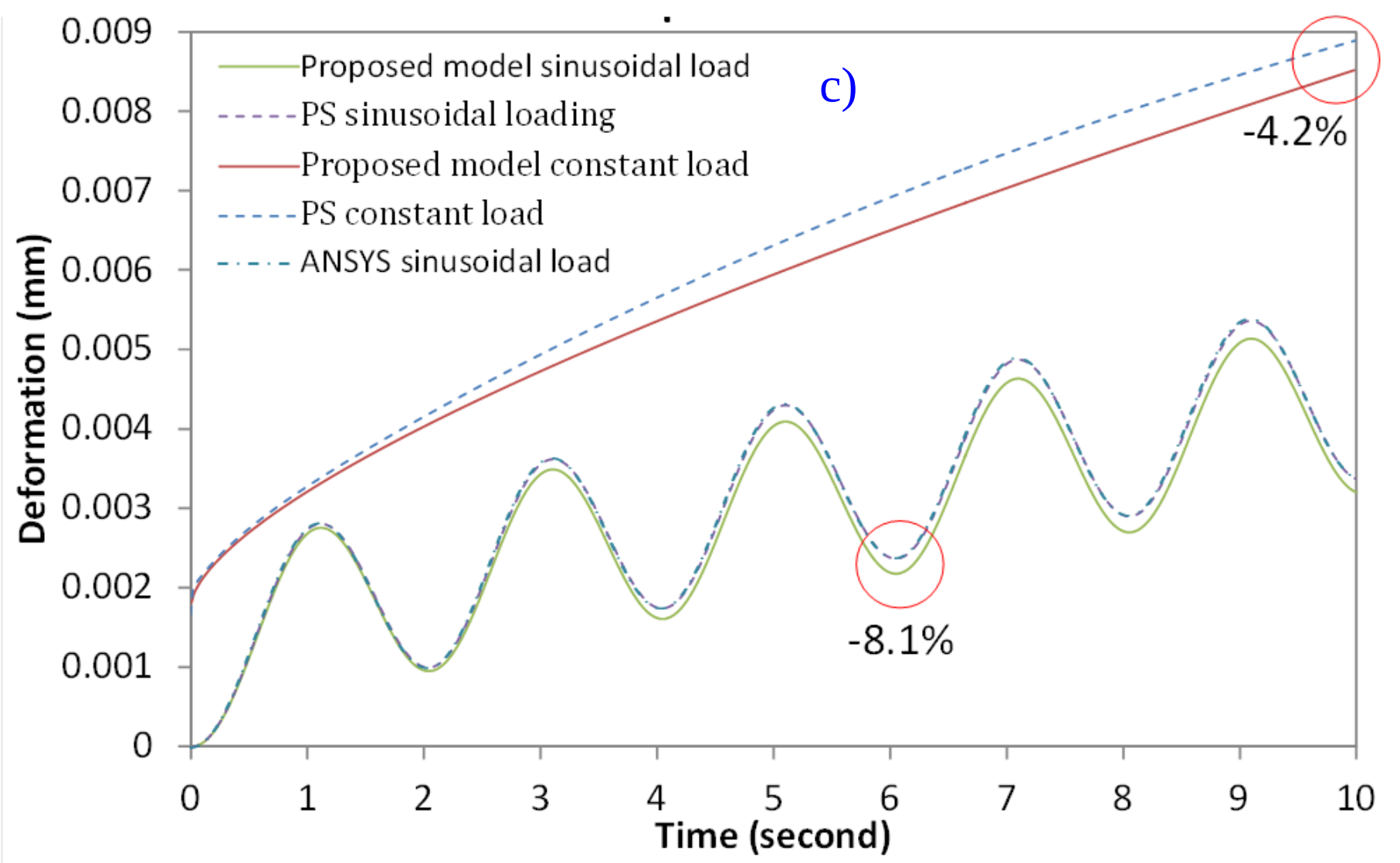

**Figure 8.** Simulated creep deformations: a) deformations under Heaviside constant loading using variable time steps: a1- results of proposed model and a2 - results of Prony series both converged at the time step of 0.01s; b) deformations under sinusoidal loading: b1- results of proposed model and b2-results of Prony series both converged at the time step length of 0.1 s and c) deformation comparison of proposed model to that of Prony series with time step length of 0.01s showing notable difference at some time points (note: the ANSYS simulation under sinusoidal loading is to validate the developed FE solution using PS model, which attains the almost identical results as that of the prosed method although using different numerical algorithm).

**B. Simulate dynamic viscoelastic stress-strain of heterogeneous structures**

To further evaluate the proposed model and its numerical accuracy, we implemented the model for modeling the stress-displacement or strain behaviors of infrastructures materials, synthetic spider silk and hydrogel as illustrated in **Figure 9**. To consider the complex condition, we firstly modeled a multilayered pavement structure on the roadway with heterogeneous material properties (Figure 9a). A dynamic loading pulse was applied on the surface of the layered structure on a circular area with a radius of 15 cm to emulate the vehicle loading. Dynamic modeling can capture the time offsets under the wave propagation effects[4165]. The top layer of AC material has been considered viscoelastic under a short time loading period[3241]. The AC material model parameters are fitted from experimental data as already presented in Figure 4. The underlying layers of base and soil can be considered elastic and their Young's moduli and Poisson's ratios are from literature[45]. The model domain is simulated by an axial symmetric FE model. Figure 9a1 presents the information of loading pattern, roadway structure, material properties, and simulated dynamic-viscoelastic vertical displacements at the measuring distances of 0, 50, and 100 cm to the loading center. Results have clearly shown that displacements have time lags to the loading pulse, and these lags increase with measuring distances due to the wave propagation effect. Some simulated displacements remain positive after the loading drops to zero and even becomes negative, which is caused by the creep behavior of AC material. Figure 9a2 presents the simulated vertical displacement-force loop of the dynamic viscoelastic model versus that of the static viscoelastic and dynamic elastic models. The loading-displacement hysteresis loop shows that the viscoelastic model (both static and dynamic) is able to simulate the stress recovery and energy dissipation after unloading. The size of loop area reduces with the increase of measuring distance. This is caused by the decay of the viscoelastic effect of the top AC layer at further fields where deformations are more dependent on the material properties of the underlying layers being considered elastic. The dynamic model induces higher energy dissipation as shown by the larger hysteresis loop (e.g. at the zero distance) than that of the static model due to its kinetic energy contribution. These results clearly show that the proposed model can reasonably simulate the dynamic and viscoelastic material behaviors of heterogeneous structures at complex loading and boundary conditions.

Figure 9b present the modeled stress-strain relationship of a spider-inspired silk threads glued to a glass substrate with the glycoprotein glue under peeling forces versus experimental measurements (laboratory testing data reproduced from[66] with permission). Results have shown the model is able to capture the nonlinear strain hardening behavior under large strain. Figure 9c presents the modeled stress-strain relationship of the polyampholyte-formed hydrogel material (laboratory data reproduced from[5] with permission). This hydrogel performs viscoelastic behaviors with stress recovery for strain up to 700%[5] and is tough to be used as a structural material. Results show that the model is able to accurately model the complex and nonlinear material behaviors with stress "softening" and then hardening. One shall note that the materials may behavior plastic deformation and fractures at a larger strain range such as for the spider or inspired silk 6. However, the proposed model in this research has only accounted for the viscoelastic part, such as for the hydrogel material studied here with a strain hardening up to 700% with stress recovery 5.

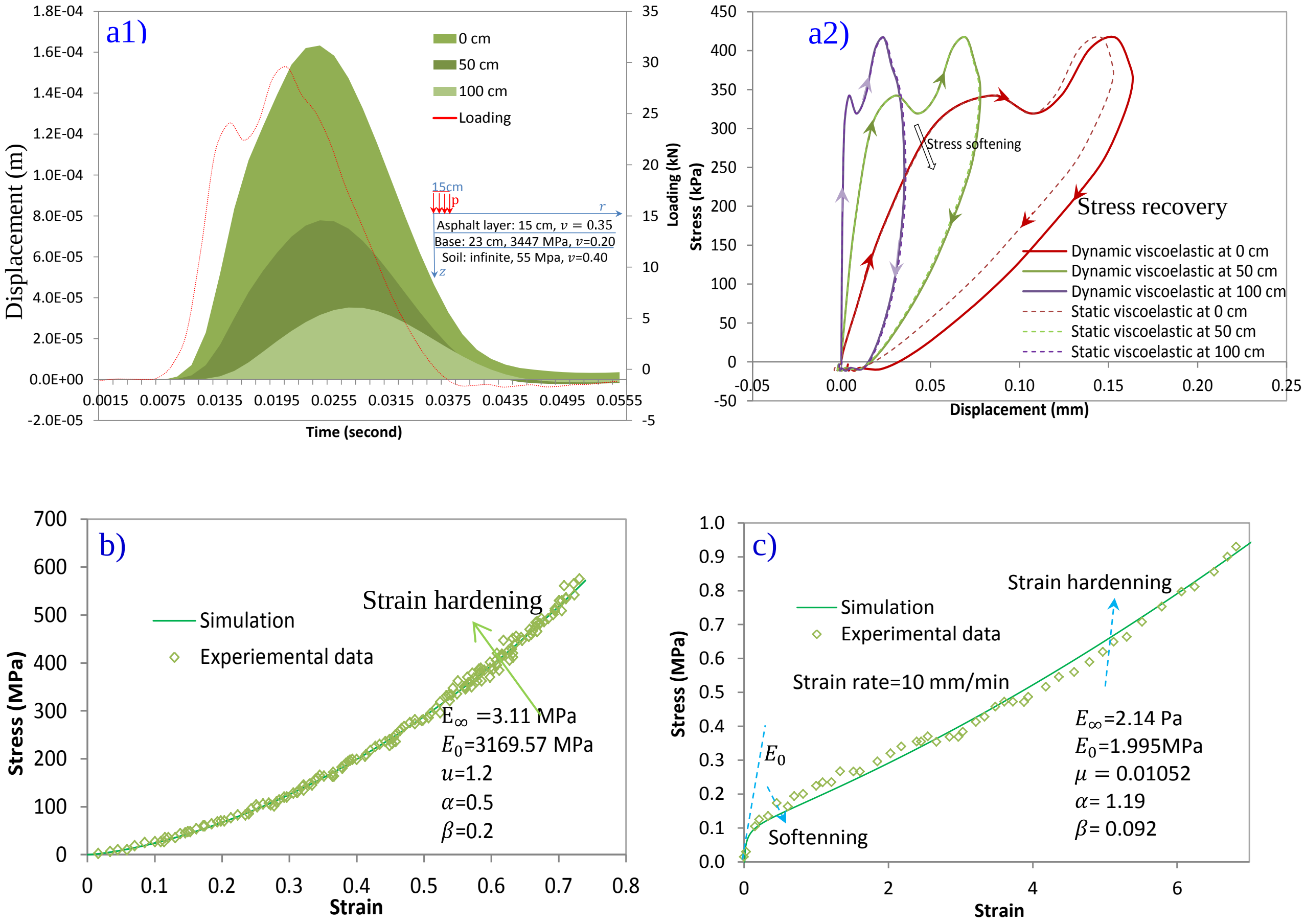

**Figure 9.** Simulated stress-strain or displacement behaviors: a) deformation of multilayered pavement structure under dynamic loading: a1. dynamic loading and simulated vertical displacements of variable distances on surface showing the time lags of displacements to loading, a2. Stress-displacement hysteresis loop showing the stress recovery and energy dissipation during unloading; b) stress-strain behavior of spider-inspired synthetic silk material adhered to steel surface showing nonlinear strain hardening behavior of modeling results vs. experimental values and c) nonlinear stress -strain behavior of the viscoelastic hydrogel material showing nonlinear stress softening and then hardening.

## VI. CONCLUSION AND FUTURE WORK

We developed a new material viscoelastic model and mathematical solution to simulate relaxation modulus and viscoelastic responses with relatively simple format for simulating a broad range of materials including the advanced and bio-materials. The physical mechanism can be interpreted by a spring network-viscous medium system with five model parameters to represent general materials considering nonlinear strain hardening We also developed a Galerkin finite-element method and robust numerical algorithms to implement the new model for dynamic analysis. We have validated the model on both experimental data and numerical simulations for different materials including natural asphalt, shape-memory polymer, spider-inspired silk, agar, and bone. The main findings can be drawn in the following:

- By satisfying the Calusius-Duhem inequality for thermodynamics consistency, the model is able to simulate reasonable creep and sinusoidal responses and energy dissipation under dynamic loading for heterogeneous materials and structures.
- The new model has a few superior features than PS - the most general model often with a large number of model parameters being used in the current state: a) it improves accuracy in predicting relaxation modulus outside of the experimental range to provide a useful tool for numerical simulations, b) it improves numerical accuracy in simulating responses of materials with competitive numerical stability, computation speed for convergence and c) it is able to simulate both the linear and nonlinear viscoelastic behavior of complex or advanced material systems;

Therefore, the model may be used as an alternative to simulate the viscoelastic behaviors of a broad range of materials including the multifunctional, advanced, and biomaterials at both experimental and numerical scales with improving accuracy while less complexity. However, as a research in progress for the future work, further tests and/or improvements of the model for a broad range of materials and cases including more variable and complex stress-strain behaviors could be performed. It would also be beneficial to develop a frequency-domain solution of the prosed model for numerical simulations. The model could be extended from the viscoelasticity to a damage model considering viscoelastic/plastic & fractures of the full strain range.

## ACKNOWLEDGEMENT

We would appreciate the support of laboratory and filed tests by the US Department of Transportation and courtesy of partial experimental data with permissions from the authors and/or publishers[5,8,58,59,66]. The first author completed the main work at UT Austin before he moved to Minnesota in May 2015 as a computational mechanics engineer.

## APPENDIX A: MATHEMATICAL PROOF FOR THE TWO-ELEMENT SYSTEM MODEL

The viscosity of the dashpot is a nonlinear function of time as follows:

$$\eta(t) = \frac{1+\mu^{\alpha} e^{\alpha t}}{\alpha \mu^{\alpha} e^{\alpha t}} \tag{21}$$

Applying a constant strain $\varepsilon$ on the two-element system (see Figure 2g), the stress equilibrium satisfies the following:

$$\varepsilon = \varepsilon_1(t) + \varepsilon_2(t) \tag{22}$$

$$\sigma(t) = E_{\infty}\varepsilon + \Delta E \varepsilon_1 (1+\varepsilon)^{\beta} \tag{23}$$

$$\Delta E \varepsilon_1 = \eta(t) \dot{\varepsilon}_2(t) \tag{24}$$

where $\varepsilon_0$ is the constant total strain of the model which is the same as that posed by the $E_{\infty}$ spring, $\varepsilon_1$ and $\varepsilon_2$ are strain of the nonlinear spring $\Delta E$ and dashpot $\eta(t)$, respectively.
Substitute equation (21) and (22) into (24), $\varepsilon_1$ can be derived as follows in sequence:

$$\frac{\Delta E \alpha \mu^{\alpha} e^{\alpha t}}{1+\mu^{\alpha} e^{\alpha t}} dt = \frac{d\varepsilon_2(t)}{\varepsilon_1(t)} = \frac{d\left(\varepsilon - \varepsilon_1(t)\right)}{\varepsilon_1(t)} = -\frac{d\varepsilon_1(t)}{\varepsilon_1(t)} \tag{25}$$

$$\varepsilon_1(\mathrm{t}) = \frac{\varepsilon}{1+\mu^{\alpha} e^{\alpha t}} \tag{26}$$

Substitute equation (26) into equation (23), the equivalent formula of $E(t)$ can be derived as follows:

$$E(t) = \sigma(t)/\varepsilon = E_{\infty} + \frac{\Delta E}{1+\mu^{\alpha} e^{\alpha t}} (1+\varepsilon)^{\beta} \tag{27}$$

## APPENDIX B: MATHEMATICAL PROOF FOR CALUSIUS-DUHEM INEQUALITY

To ensure the thermodynamic consistency, the Calusius-Duhem inequality shall be satisfied 54 as follows:

$$\rho T \gamma = -\rho \dot{\phi} + \sigma \dot{\varepsilon} - \rho s \dot{T} - \frac{q}{T} \frac{\partial T}{\partial x} \geq 0 \tag{28}$$

where $\rho \theta \gamma$ is a specific energy dissipation term, $\phi$ is the specific free energy, $\varepsilon$ is the total stain of the system, and $s$ is the specific entropy.
The total strain can be decomposed into two parts as follows:

$$\varepsilon = \varepsilon_m + \varepsilon_T \tag{29}$$

where $\varepsilon_m$ is the mechanical strain and $\varepsilon_T$ is the thermal strain due to thermal expansion or contraction.
The mechanical strain can be further decomposed into two parts as follows:

$$\varepsilon_m = \varepsilon_{\mathrm{e}} + \varepsilon_{\mathrm{v}} \tag{30}$$

where $\varepsilon_e$ is the elastic strain posed by the spring of $(E_0 - E_{\infty})$, and $\varepsilon_{\mathrm{v}}$ is inelastic strain posed by the dashpot.
The thermal strain is expressed as follows:

$$\varepsilon_T = \alpha_v (T - T_0) \tag{31}$$

where $\alpha_v$ is the coefficient of thermal expansion or contraction, $T$ is the current temperature, and $T_0$ is the reference temperature.
The total stress can be decomposed to two parts as well:

$$\sigma = \sigma_{\mathrm{e}} + \sigma_{\mathrm{vE}} \tag{32}$$

where $\sigma_e$ is an elastic stress posed by the elastic spring of $E_{\infty}$, and $\sigma_{\mathrm{vE}}$ is the viscoelastic stress posed by the spring-dashpot series (shown in Figure 2g).
The free energy of the model system is given as follows:

$$\phi = \phi_e(\varepsilon_m, T) + \phi_v(\varepsilon_{\mathrm{e}}, T) + \xi(T) \tag{33}$$

$\phi_e(\varepsilon_m, T)$ is the energy stored in the spring of $E_\infty$, $\phi_v(\varepsilon_\mathrm{e}, T)$ is the energy stored in the spring-dashpot series, $\xi(T)$ is an energy related to the heat capacity.
Submit Equations (29) and (33) into Equation (28) to reach the following equilibrium:

$$\rho T\gamma = -\rho\left(\dot{\phi}_e(\varepsilon_m, T) + \dot{\phi}_v(\varepsilon_\mathrm{e}, T) + \dot{\xi}(T)\right)$$

$$+\sigma(\dot{\varepsilon}_m + \dot{\varepsilon}_T) - \rho s\dot{T} - \frac{q}{T}\frac{\partial T}{\partial x} \quad (34)$$

Substitute Equations (30) to (32) into Equation (34), and apply the chain rule to achieve the following:

$$\rho T\gamma = -\rho\left(\frac{\partial\phi_e}{\partial\varepsilon_m}\dot{\varepsilon}_m + \frac{\partial\phi_e}{\partial T}\dot{T} + \frac{\partial\phi_v}{\partial\varepsilon_\mathrm{e}}\dot{\varepsilon}_\mathrm{e} + \frac{\partial\phi_v}{\partial T}\dot{T} + \frac{\partial\xi}{\partial T}\dot{T}\right)$$

$$+\left(\sigma_\mathrm{e}\dot{\varepsilon}_m + \sigma_\mathrm{v}(\dot{\varepsilon}_\mathrm{e} + \dot{\varepsilon}_\mathrm{v}) + \alpha\sigma\dot{T}\right) - \rho s\dot{T} - \frac{q}{T}\frac{\partial T}{\partial x} \quad (35)$$

This can ber re-arranged as follows:

$$\rho T\gamma = \left(\sigma_\mathrm{e} - \rho\frac{\partial\phi_e}{\partial\varepsilon_m}\right)\dot{\varepsilon}_m + \left(\sigma_\mathrm{vE} - \rho\frac{\partial\phi_v}{\partial\varepsilon_\mathrm{e}}\right)\dot{\varepsilon}_\mathrm{e}$$

$$+\left[\alpha\sigma + \frac{\partial\phi_e}{\partial T} - \rho\left(s + \frac{\partial\phi_v}{\partial T}\right) + \frac{\partial\xi}{\partial T}\right]\dot{T} + \sigma_\mathrm{v}\dot{\varepsilon}_\mathrm{v} - \frac{q}{T}\frac{\partial T}{\partial x} \quad (36)$$

To satisfy $\rho T\gamma \geq 0\ \forall\ \dot{\varepsilon}_m, \dot{\varepsilon}_e, \dot{T}$, the coefficients of the terms $(\dot{\varepsilon}_m, \dot{\varepsilon}_e, \dot{T})$ have to vanish, and thus:

$$\sigma - \rho\frac{\partial\phi_e}{\partial\varepsilon_m} = 0 \quad (37a)$$

$$\sigma_\mathrm{vE} - \rho\frac{\partial\phi_v}{\partial\varepsilon_\mathrm{e}} = 0 \quad (37b)$$

$$\alpha\sigma + \frac{\partial\phi_e}{\partial T} - \rho\left(s + \frac{\partial\phi_v}{\partial T}\right) + \frac{\partial\xi}{\partial T} = 0 \quad (37c)$$

The viscoelastic strain rate is:

$$\dot{\varepsilon}_\mathrm{v} = \frac{\sigma_\mathrm{v}}{\eta} \quad (38)$$

According to the Fourier's rule, the heat flux is expressed as follows:

$$q = -\gamma\frac{\partial T}{\partial x} \quad (39)$$

Submit equations (37) to (39) into Equation (28):

$$\rho T\gamma = \sigma_{\mathrm{v}}\dot{\varepsilon}_{\mathrm{v}} - \frac{q}{T}\frac{\partial T}{\partial x} = \frac{\sigma_{\mathrm{v}}^2}{\eta} + \gamma\left(\frac{\partial T}{\partial x}\right)^2 \geq 0 \tag{40}$$

Therefore, $\rho T\gamma \geq 0$.

## APPENDIX C: NUMERICAL METHOD AND PROCEDURE FOR SOLVING PDES

For the 3-D model domain, the stress-displacement relationship of the viscoelastic solid can be expressed as follows:

$$\boldsymbol{\sigma}(t) = 2\int_0^t G(t-\tau,\varepsilon)\frac{\partial\left(\frac{1}{2}[\nabla\boldsymbol{u}(\tau)+\nabla\boldsymbol{u}(\tau)^{\boldsymbol{T}}]-\frac{1}{3}\nabla\cdot\boldsymbol{u}(\tau)\right)}{\partial\tau}d\tau$$

$$+3\mathbf{I}\int_0^t K(t-\tau,\varepsilon)\frac{\partial\left(\frac{1}{3}\nabla\cdot\boldsymbol{u}(\tau)\right)}{\partial\tau}d\tau \tag{41}$$

where the first part is the deviatoric stress tensor, and the second term is the hydrostatic stress.

### C.1 Galerkin Formulation

Multiply a test function $\boldsymbol{p}(t)$ on both sides of the strong form (Equation (18)), and then integrate with the space and time domains to form a weak form as follows:

$$\int_0^{t_d}\int_\Omega(\nabla\cdot\boldsymbol{\sigma})\cdot\boldsymbol{p}(t)d\Omega dt + \int_0^{t_d}\int_\Omega b\cdot\boldsymbol{p}(t)d\Omega dtt - \int_0^{t_d}\int_\Omega\rho\ddot{\boldsymbol{u}}(t)\ \cdot\boldsymbol{p}(t)d\Omega dt = 0 \tag{42}$$

According to divergence theory, the divergence of stress can be decomposed as two parts:

$$(\nabla\cdot\boldsymbol{\sigma})\cdot\boldsymbol{p}(t)d\Omega =\ (\boldsymbol{\sigma}\cdot nds)\cdot\boldsymbol{p}(t) - \boldsymbol{\sigma}{:}\,\nabla\boldsymbol{p}(t) \tag{43}$$

where $(\boldsymbol{\sigma}\cdot nds)$ is the surface trace as equivalient to exterial loading of $f(t)$ applied on the surface of the model domain. Substitute Equation (43) into (42), the weak form can be re-expressed as follows:

$$\int_\Omega\boldsymbol{\sigma} : \nabla\boldsymbol{p}(t)d\Omega - \int_\Omega b\cdot\boldsymbol{p}(\boldsymbol{t})d\Omega dt + \int_\Omega\rho\ddot{\boldsymbol{u}}(t)\ \cdot\boldsymbol{p}(t)d\Omega - \int_{\partial\Omega}f(t)\cdot\boldsymbol{p}(t)dsdt = 0 \tag{44}$$

Subsitute Equation (41) into Equation (44) to reach the final weak form in a continum state:

$$\int_\Omega R(t-\tau,\varepsilon){:}\,\boldsymbol{\nabla}\boldsymbol{p}(t)d\Omega + \int_\Omega\rho\ddot{u}(t)\cdot\boldsymbol{p}(t)d\Omega = \left[\int_{\partial\Omega}f(t)\cdot\boldsymbol{p}(t)ds + \int_\Omega b\cdot\boldsymbol{p}(t)d\Omega\right] \tag{45}$$

where $R(t-\tau)$ is the relaxation modulus in the 3-D domain defined as follows:

$$R(t-\tau) \coloneqq 2\int_0^t G(t-\tau,\varepsilon)\left(\frac{1}{2}[\nabla\dot{\boldsymbol{u}}(\tau)+\nabla\dot{\boldsymbol{u}}(\tau)^{\boldsymbol{T}}] - \frac{1}{3}\nabla\cdot\boldsymbol{u}(\tau)\right)d\tau$$

$$+3\mathbf{I}\int_0^t K(t-\tau,\varepsilon)\left(\frac{1}{3}\boldsymbol{\nabla}\cdot\dot{\boldsymbol{u}}(\tau)\right)d\tau \tag{46}$$

The displacement and test function are then discretized to that at the FE nodes as follows:

$$\boldsymbol{u}(t) = \sum u_i(t)N_i \tag{47}$$

$$\boldsymbol{p}(t) = \sum p_i(t)N_i \tag{48}$$

where $N_i$ is the shape function of the $i^{th}$ FE node. Substitute the discretized form of $u$ and $p$ into Equation (45):

$$\int_0^{t_d}\int_\Omega\int_0^t \boldsymbol{B^T}R(t-\tau,\varepsilon)\boldsymbol{B}\dot{u}(\tau)d\tau\, p(t)d\Omega\, dt + \int_0^{t_d}\int_\Omega \boldsymbol{N^T}\rho\boldsymbol{N}\ddot{u}(t)p(t)d\Omega\, dt$$

$$= \left[\int_0^{t_d}\int_{\partial\Omega}\boldsymbol{N_\Gamma^T}f(t)p(t)ds\, dt + \int_0^{t_d}\int_\Omega \boldsymbol{N^T}bp(t)d\Omega\, dt\right] \tag{49}$$

where $\boldsymbol{N}$ is the shape function matrix consisting of all $\phi_i$, $\boldsymbol{B} = \nabla\boldsymbol{N}$ is the strain-displacement matrix, $\boldsymbol{N_\Gamma^T}$ is the transposed shape function matrix for the 2-D loading area.
$p(t)$ is arbitrary and can be dismissed on both sides of the equation. Thus, the following weak form shall satisfy $\forall t \in [0, t_d]$:

$$\int_\Omega\int_0^t \boldsymbol{B^T}R(t-\tau,\varepsilon)\boldsymbol{B}\dot{u}(\tau)d\tau\, d\Omega + \left(\int_\Omega \boldsymbol{N^T}\rho\boldsymbol{N}d\Omega\right)\ddot{u}(t)$$

$$= \left[\int_{\partial\Omega}\boldsymbol{N_\Gamma}^{\boldsymbol{T}}f(t)ds + \int_\Omega \boldsymbol{N^T}bd\Omega\right] \tag{50}$$

Let:

$$\boldsymbol{R}(t-\tau,\varepsilon) = \int_\Omega \boldsymbol{B^T}R(t-\tau,\varepsilon)\boldsymbol{B}d\Omega$$

$$= \boldsymbol{R}'(t-\tau)\ (1+\mu)^\alpha \tag{51}$$

where $R'(t-\tau)$ is the relaxation modulus matrix excluding the influence of strain function.
This weak form can be reduced to a simplified format as follows:

$$\int_0^t \boldsymbol{R}'(t-\tau)\dot{u}(\tau)(1+u)^a\, d\tau + \mathbf{M}\ddot{u}(t) = \mathfrak{R} \tag{52}$$

where $\mathbf{M} = \int_\Omega \boldsymbol{N^T}\rho\boldsymbol{N}\mathrm{d}\Omega$ as a mass matrix, and $\mathfrak{R} = \int_{\partial\Omega_4}\boldsymbol{N_\Gamma}^{\boldsymbol{T}}f(t)ds + \int_\Omega \boldsymbol{N^T}b\mathrm{d}\Omega$ as the load vector.

**C.2 Discretization and Solution**

The time doamin $t \in [t_0, t_d]$ is discretized to $N$ time steps for $k = 1,2 \ldots N$, and for each of them the sub-time domain $\tau \in [0, t]$ includes $k$ time steps for $j = 1,2 \ldots k$. Thus, Equation (52) can be rewritten as follows after discretizing $\tau \in [0, t]$ to $k$ time steps:

$$\sum_{j=\mathbf{1}}^k \int_{t_{j-1}}^{t_j} \boldsymbol{R}'(t-\tau)\dot{u}(\tau)(1+u)^\beta d\tau + \mathbf{M}\ddot{u}(k) = \mathfrak{R} \tag{53}$$

$$\sum_{j=\mathbf{1}}^{k} \int_{t_{j-1}}^{t_j} \boldsymbol{R}'(t-\tau)\dot{u}(\tau)(1+u)^{\beta} d\tau = \Re \tag{54}$$

The first order gradient or velocity term $\dot{u}(\tau)$ at the sub-time of $\tau \in [t_{j-1}, t_j]$ in this weak form can be discretized as follows according to the explicit Euler's rule:

$$\dot{u}(\tau) = \frac{[u(j)-u(j-1)]}{\Delta t} \tag{55}$$

where $u(\tau)$ is displacement at the sub-time $\tau \in [t_{j-1}, t_j]$ for $j = 1,2 \ldots k$.
This discretized velocity at the sub-time discretization can be taken out of the time integration of $\tau \in [t_{j-1}, t_j]$, and thus Equation (53) can be re-expressed as follows:

$$\sum_{j=\mathbf{1}}^{k} \boldsymbol{J}(j)\left[u_{\mathrm{j}} - u_{\mathrm{j-1}}\right](1+\mu_j)^{\beta} + \mathbf{M}\ddot{u}_k = \Re \tag{56}$$

where $J(j)$ is a viscoelastic stiffness matrix at the sub-time step for $j = 1,2,3 \ldots\ldots k$ and is defined as follows for computation purpose:

$$\boldsymbol{J}(j) = \int_{\mathrm{t_{j-1}}}^{t_j} \boldsymbol{R}'(t-\tau)d\tau \tag{57}$$

Here material's possion's ratio $v$ is considered constant at constant temperatures, the relationship between Young's and shear and bulk relaxation modulus can be attained from the elastic model constitutive relationship. Therefore, $J(t)$ can be rederived as follows:

$$\boldsymbol{J}(j) = \mathbf{C_e} \int_{t_{j-1}}^{t_j} E(t-\tau)d\tau \tag{58}$$

where $\mathbf{C_e}$ is an elastic matrix such that:

$$\mathbf{C_e} = \int_{\Omega} \boldsymbol{B^T} \mathbb{C} \boldsymbol{B} d\Omega \tag{59}$$

where $\mathbb{C}$ is the 4$^{\text{th}}$ order elasticity tensor over elastic modulus E.
For the proposed model, substitute Equation (9) ($\tau_0$=1) into Equation (58) the viscoelastic stiffness matrix of the proposed model $\boldsymbol{J}_P(j)$ can be calculated as follows:

$$\boldsymbol{J_P}(j) = \int_{\Omega} \int_{t_{j-1}}^{t_j} \boldsymbol{B^T}\left(E_\infty + (E_0 - E_\infty)\frac{1}{1+\mu e^{\alpha \log(t-\tau)}}\right)\mathbb{C}\boldsymbol{B} d\tau d\Omega \tag{60}$$

$\boldsymbol{J_P}(j)$ can be discretized as follows following the trapzoidal rule:

$$\boldsymbol{J_P}(j) = \mathbf{C_e}\left\{E_\infty + \frac{(E_0 - E_\infty)}{2}\left[\frac{1}{1+\mu e^{\alpha \log\left(t-t_j\right)}} + \frac{1}{1+\mu e^{\alpha \log\left(t-t_{j-1}\right)}}\right]\right\}\Delta t \tag{61}$$

where $t_{r,j}$ and $t_{r,j-1}$ are the $j^{th}$ and $(j-1)^{th}$ time, $\Delta t = t_j - t_{j-1}$.
For the PS model, substitute Equation (1) into Equation (58) $J(j)$ can be derived as follows:

$$\boldsymbol{J_{PS}}(j) = \mathbf{C_e} \int_{t_{j-1}}^{t_j} \left[E_\infty + \sum_i^N E_i e^{-\frac{E_i}{\eta_i}(t-\tau)}\right] d\tau \tag{62}$$

With the integration of sub-time step $[t_{j-1}, t_j]$, $\boldsymbol{J}(j)$ can be re-derived as follows:

$$\boldsymbol{J_{PS}}(j) = \mathbf{C_e}\left\{E_\infty \Delta t + \sum_i^N \eta_i \left[e^{-\frac{E_i}{\eta_i}(t-t_j)} - e^{-\frac{E_i}{\eta_i}(t-t_{j-1})}\right]\right\} \tag{63}$$

Different algorithms have been developed for the time discretization of acceleration, among which the Houbolt method is less dependent of time step and unconditional stable 63. Therefore it is chosen herein for discretizing the acceleration as follows:

$$\ddot{u}_k = [2u_k - 5u_{k-1} + 4u_{k-2} - u_{k-3}]/\Delta t^2 \tag{64}$$

where $u(k)$ is displacement, $\Delta t$ is time step length. As simulation results, a relatively great time step length (i.e. 0.001 second) can be used to achieve accurate numerical simulations based on the sensitivity analysis, which improves computation speed.

For solving the nonlinear viscoelastic strain, substitute Equation (64) into Equation (56) to attain the final weak form with discretized time as follows:

$$\Re = \sum_{j=\mathbf{1}}^{k} \boldsymbol{J}(j)[u(j) - u(j-1)]\big(1 + u(j)\big)^{\beta}$$

$$+[2u(k) - 5u(k-1) + 4u(k-2) - u(k-3)]\frac{\mathbf{M}}{\Delta t^2} \tag{65}$$

where $u(k)$ is displacement at the $k^{th}$ time step for $k = 1,2,3 \ldots n$; $u(j)$ is displacement at the $j^{th}$ sub-time step or time $t_j$ for $j = 1,2,3 \ldots k$.

$$\boldsymbol{J}(k)[u_{\mathrm{k}} - u_{\mathrm{k-1}}](1 + \mu_k)^{\alpha+1} + \sum_{j=\mathbf{1}}^{k-1} \boldsymbol{J}(j)\big[u_{\mathrm{j}} - u_{\mathrm{j-1}}\big]\big(1 + \mu_j\big)^{\beta}$$

$$+[2u_{\mathrm{k}} - 5u_{\mathrm{k-1}} + 4u_{\mathrm{k-2}} - u_{\mathrm{k-3}}]\frac{\mathbf{M}}{\Delta t^2} - \Re = 0 \tag{66}$$

Let the above as a equation $f(u_0, u_1, \ldots \ldots u_k)$=0, Newton's iteration method can be applied to solver the above equation to determine $\mu(k)$ by calculating the tangent at each iteration step as:

$$\frac{\partial f}{\partial u_k} = J(k)[1 + u_k]^{\beta} + J(k)[u_k - u_{k-1}]\beta(1 + u_k)^{\beta-1} + \frac{2M}{\Delta t^2} \tag{67}$$

The stress can be solved as follows:

Stress analysis method:

$$\sigma(t) = \int_0^t \boldsymbol{R}(t-\tau, \varepsilon)d\varepsilon = \int_0^t \boldsymbol{R}'(t-\tau)(1+u)^{\beta} d\varepsilon \tag{68}$$

For the linear viscoelasticity problem, let $\beta = 0$ in Equation (65):

$$\Re = \sum_{j=1}^{k}[u(j) - u(j-1)]\frac{J(j)}{\Delta t}$$

$$+[2u(k) - 5u(k-1) + 4u(k-2) - u(k-3)]\frac{\mathbf{M}}{\Delta t^2} \tag{69}$$

where $u(k)$ is displacement at the $k^{th}$ time step for $k = 1,2,3 \dots n$; $u(j)$ is displacement at the $j^{th}$ sub-time step or time $t_j$ for $j = 1,2,3 \dots k$.

Equation (69) can be re-arranged and reduced to a linear system as follows:

$$\mathbf{K_{dve}}u(k) = \mathfrak{R}^* \tag{70}$$

$$\mathbf{K_{dve}} = \frac{2\mathbf{M} + \boldsymbol{J}(k)\Delta t}{\Delta t^2} \tag{71}$$

$$\mathfrak{R}^* = \mathfrak{R} + \left[\frac{\boldsymbol{J}(k) - \boldsymbol{J}(k-1)}{\Delta t} + \frac{5\mathrm{M}}{\Delta t^2}\right]u(k-1)$$

$$+\left[\frac{\boldsymbol{J}(k-1) - \boldsymbol{J}(k-2)}{\Delta t} - \frac{4\mathbf{M}}{\Delta t^2}\right]u(k-2)$$

$$+\left[\frac{\boldsymbol{J}(k-2) - \boldsymbol{J}(k-3)}{\Delta t} + \frac{\mathbf{M}}{\Delta t^2}\right]u(k-3)$$

$$-\sum_{j=1}^{k-4}\frac{\boldsymbol{J}(j)}{\Delta t}u(\mathrm{j}) + \sum_{j=1}^{k-3}\frac{\boldsymbol{J}(j)}{\Delta t}u(\mathrm{j}-1) \tag{72}$$

where $\mathbf{K_{dve}}$ is the dynamic viscoelastic stiffness matrix and $\mathfrak{R}^*$ is the dynamic viscoelastic "loading" vector.

We adopt the factorization method with banded matrix storage methodology for solving the global linear system in Equation (70). Since $\mathbf{K_{dve}}$ is symmetric positive definite, it can be stored in a banded matrix to save computation memory and time. Consequently, the banded $\mathbf{K_{dve}}$ is decomposed to the upper and lower triangular matrices ($\mathbf{K}_l$ and $\mathbf{K}_u$) as follows:

$$\mathbf{K_{dve}} = \mathbf{K}_l\mathbf{K}_u \tag{73}$$

where $\mathbf{K}_u$ matrix is equivalent to the conjugate transpose matrix of $\mathbf{K}_l$. Substitute Equation (73) into Equation (70), the global linear system can be re-written as follows:

$$\mathbf{K}_l\{\mathbf{K}_u u(k)\} = \mathfrak{R}^* \tag{74}$$

The solution of this linear system involves two steps in sequence. Firstly one calculates a vector term ($q$):

$$q = \mathbf{K}_u u(k) = \mathbf{K}_l^{-1}\mathfrak{R}^* \tag{75}$$

Secondly, one solves the displacement $u$ from the following equilibrium:

$$u(k) = \mathbf{K}_u^{-1}q = \mathbf{K}_u^{-1}\{\mathbf{K}_u u(k)\} \tag{76}$$

Given the initial condition, boundary conditions, and the loading function at each time step, the displacement vector $u(t)$ at each time step can be calculated by one solution of this linear system. The time domain $t \in [t_0, t_d]$ is discretized to $N$ finite steps. The forward computation starts at zero and ends at $t_d$ with total $N$ time steps. The response calculated at the current time $t$ and step $k$ is dependent on those determined at previous time steps for $j = 1,2,3 \ldots \ldots k-1$. The $N$ total time steps require total $N$ solutions of linear system.

At time zero with zero loading, there are neither displacement ($u(0) = 0$) value, nor acceleration ($\ddot{u}(0) = 0$). Following the Houbolt method the calculation of $u(1)$ requires inputs of $u(-2)$ and $u(-1)$ which are unknown. Therefore, the central and backward finite different method (BFD/CFD) is adopted here to solve $u(1)$ and $u(-1)$. Following the Central CFD, the acceleration at zero time is discretized as follows:

$$\partial^2 u/\partial t^2|_{\mathrm{t=0}} = [u(1) - 2\mathrm{u}(0) + u(-1)]/\Delta \mathrm{t}^2 = 0 \tag{77}$$

At the zero time step the lienar system of the governing state equation satisfies:

$$\boldsymbol{J}(0)[u(0) - u(-1)]/\Delta t + \mathbf{M}\ddot{u}(0) = \Re(0) \tag{78}$$

From Equations (77) and (78) the followings satisfy:

$$\boldsymbol{J}(0)u(-1) = -\Delta t \Re(0) \tag{79}$$

$$u(1) = -\mathrm{u}(-1) \tag{80}$$

Thus $u(-1)$ and $u(1)$ can be solved. Given $u(0) = 0, u(2)$ is solved following the Houbolt time discretization, and so on for $u(3), u(4) \ldots u(N)$. With displacements calculated at all FE nodes, the strain and then stress responses can be solved following the deformation-strain-stress constitutive relationships, which are not discussed in this research.

## APPENDIX D: MATHEMATICAL DERIVATION FOR THE ERROR OF VISCOELASTIC STIFFNESS MATRIX

Here $f(\xi) = R(t-\tau)$ for the 3-D model domain and $f(\xi) = E(t-\tau)$ for the 1-D model domain. The numerical error for $J_J$ is derived as follows:

$$\mathrm{Err_J} = \frac{(t_2 - t_1)^3}{12N^2} J''(\xi) = -\frac{\Delta t^3}{12} \mathbf{C_e} \frac{\partial^2 E(t-\xi)}{\partial \xi^2} \quad \forall \xi \in [t_1, t_2] \tag{81}$$

where $\mathbf{C}_\mathrm{e}$ is an elastic matrix. The proposed model formula in Equation (7) can be rewritten as follows:

$$E(t-\xi) = 10^{E_\infty + \frac{\Delta E}{1 + A(t-\xi)^\gamma}} \tag{82}$$

where $A = K^\beta$ and $\gamma = -\frac{\beta}{ln10}$.

For validation purpose, the error analysis for the proposed model is compared with that of PS and thus the modulus is considered strain rate independent here. Thus, the first and second order derivative of $E(\xi)$ with respect to $\xi$ is derived as follows in consequence:

$$\frac{\partial E(t-\xi)}{\partial \xi} = \ln(10)A(\Delta E)E(t-\xi)\frac{\gamma(t-\xi)^{\gamma-1}}{(1+A(t-\xi)^{\gamma})^2} \tag{83}$$

$$\frac{\partial^2 E(t-\xi)}{\partial \xi^2} = \ln(10)A\Delta E\left\{\frac{\partial E(t-\xi)}{\partial \xi}\frac{\gamma(t-\xi)^{\gamma-1}}{[1+A(t-\xi)^{\gamma}]^2}\right.$$

$$\left.+E(t-\xi)\left[\frac{-\gamma(\gamma-1)(t-\xi)^{\gamma-2}}{[1+A(t-\xi)^{\gamma}]^2}+\frac{2A\gamma^2(t-\xi)^{2(\gamma-1)}}{[1+A(t-\xi)^{\gamma}]^3}\right]\right. \tag{84}$$

Substitute Equation (84) into Equation (19), the numerical error of $J(j)$ for the proposed model can be derived as follows:

$$\mathrm{Err_P}(t) = \frac{-\ln(10)A\Delta E\Delta \mathrm{t}^3}{12}\mathbf{C_e}\left\{\frac{\partial E(t-\xi)}{\partial \xi}\frac{\gamma(t-\xi)^{\gamma-1}}{[1+A(t-\xi)^{\gamma}]^2}\right.$$

$$\left.+E(t-\xi)\left[\frac{-\gamma(\gamma-1)(t-\xi)^{\gamma-2}}{[1+A(t-\xi)^{\gamma}]^2}+\frac{2A\gamma^2\xi^{2(\gamma-1)}}{[1+A(t-\xi)^{\gamma}]^3}\right]\right. \quad \forall\xi \in [t_{j-1}, t_j] \tag{85}$$